\pdfoutput=1

\documentclass[11pt]{article}

\usepackage[pdftex]{graphicx,color} 
\usepackage{jheppub}
\usepackage{amsmath,amsfonts,amsthm,amssymb} 
\usepackage[x11names, rgb]{xcolor}
\usepackage{empheq}
\usepackage{multirow}
\usepackage{mathtools}
\usepackage{comment}
\usepackage{dsdshorthand}
\usepackage[utf8]{inputenc}
\usepackage{bbold}
\usepackage{bbm}
\usepackage[T1]{fontenc}
\usepackage{simplewick} 
\usepackage{makecell}	
\usepackage{tabu}		
\usepackage[makeroom]{cancel} 
\usepackage[normalem]{ulem} 

\makeatletter
\newcommand{\hathat}[1]{%
	\begingroup%
	\let\macc@kerna\z@%
	\let\macc@kernb\z@%
	\let\macc@nucleus\@empty%
	\hat{\raisebox{.2ex}{\vphantom{\ensuremath{#1}}}\smash{\hat{#1}}}%
	\endgroup%
}
\makeatother

\makeatletter
\newcommand{\hathatexp}[1]{%
	\begingroup%
	\let\macc@kerna\z@%
	\let\macc@kernb\z@%
	\let\macc@nucleus\@empty%
	\hat{\raisebox{-.2ex}{\vphantom{\ensuremath{#1}}}\smash{\hat{#1}}}%
	\endgroup%
}
\makeatother

\newcommand{\bea}{\begin{equation}\begin{aligned}}
\newcommand{\eea}[1]{\label{#1}\end{aligned}\end{equation}}
\newcommand{\boa}{\begin{align}}
\newcommand{\eoa}{\end{align}}
\newcommand{\beq}{\begin{equation}}
\newcommand{\eeq}{\end{equation}}

\newcommand\eqq[1]{eq.~(\ref{eq:#1})}

\def\d{\delta}

\def\th{\theta}
\def\D{\Delta}

\def\Dh{\hat{\Delta}}

\def\f{\phi}

\def\half{\frac{1}{2}}

\def\g{\gamma}
\def\a{\alpha}
\def\e{\eta}
\def\ep{\epsilon}

\def\k{\kappa} 
\newcommand{\nn}{\nonumber\\}

\newcommand{\pa}{\partial}
\newcommand{\eq}{&\quad}

\newcommand{\rig}{\right.}
\newcommand{\lef}{\left.}
\newcommand{\lan}{\langle}
\newcommand{\ran}{\rangle}
\newcommand{\para}{\parallel}

\newcommand{\mco}{\mathcal{O}}

\newcommand{\hD}{{\hat{\Delta}}}
\newcommand{\hg}{{\hat{\gamma}}}
\newcommand{\hp}{{\hat{\phi}}}

\newcommand{\Dhh}{\hathat{\Delta}}
\newcommand{\Dhhe}{\hathatexp{\Delta}}
\newcommand{\hhg}{\hathat{\gamma}}
\newcommand{\hO}{\hat{\mathcal{O}}}
\newcommand{\hhO}{\hathat{\mathcal{O}}}

\newcommand{\vph}{\varphi}

\newcommand{\al}{\alpha}
\newcommand{\bet}{\beta}

\newcommand{\ph}{\phi}

\newcommand{\la}{\lambda}
\newcommand{\m}{\mu}
\newcommand{\n}{\nu}

\newcommand{\rh}{\rho}

\newcommand{\ps}{\psi}

\newcommand{\Ga}{\Gamma}

\newcommand{\X}{\Xi}

\preprint{UUITP-55/21}

\title{Analytic structure and conformal bootstrap for intersecting boundaries}
\title{Interacting scalar in a wedge CFT}
\title{Interacting conformal scalar in a wedge}
\author{Agnese Bissi, Parijat Dey, Jacopo Sisti, and Alexander Söderberg}

\affiliation{Department of Physics and Astronomy,
	Uppsala University,\\
	Box 516,
	SE-751 20 Uppsala,
	Sweden}
\emailAdd{agnese.bissi@physics.uu.se}\emailAdd{parijat.dey@physics.uu.se}\emailAdd{jacopo.sisti@physics.uu.se}\emailAdd{alexander.soderberg@physics.uu.se}

\makeatletter
\gdef\@fpheader{}
\makeatother

\abstract{We study a class of two-point functions in a conformal field theory near a wedge. This is a set-up with two boundaries intersecting at an angle $\theta$. We compute it as a solution to the Dyson-Schwinger equation of motion for a quartic interaction in the $d=4-\epsilon$ bulk and on one of the boundaries in $d=3-\epsilon$, up to order $\mathcal{O}(\epsilon)$.  We have extracted the anomalous dimensions from such correlators, which we complemented with Feynman diagrams computations. }

\begin{document}
\maketitle

\section{Introduction and outlook}
\label{sec:intro}

Conformal field theories (CFT's) play an important role in understanding the physics of critical phenomena. In particular, critical exponents are related to the quantum numbers of local operators in a CFT.  However, if we would like to make connection with real world systems, we cannot avoid discussing the presence of boundaries or defects. Another reason to study these instances come from string theory, where the appearance of D-branes as boundaries is central in this context. The presence of boundaries and/or defects reduces the amount of symmetry preserved, but the residual symmetry is still powerful enough to constrain the structure of correlators, making them more intricate but richer. In particular it is possible to define an additional operator product expansion (OPE), the so called bulk-to-boundary or bulk-to-defect OPE, which allow us to expand an operator in the presence of a boundary/defect as an infinite sum of boundary/defect operators. 

\begin{figure} 
	\centering
	\includegraphics[width=0.7\textwidth]{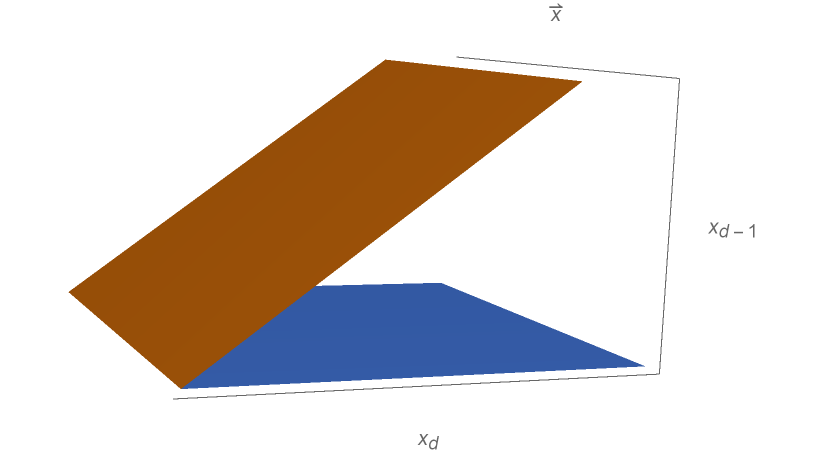}
	\caption{An illustration of the setup we consider, where the blue plane is the wall, and the orange is the ramp. The wedge is the intersection of the two boundaries.}
	\label{Fig: edge}
\end{figure}

In this paper we consider a $d$-dimensional CFT living in a wedge-shaped region bounded by two $(d-1)$-dimensional half-planes intersecting at an angle $\theta \in (0,2\pi)$, as in Fig. \ref{Fig: edge}. To be concrete, we take one half-plane, which we call \textit{wall}, located at $x_{d-1} = 0$, and the other one, the \textit{ramp}, defined by $x_{d-1} = \tan\theta \, x_d$. With a little abuse of terminology, we will call \textit{wedge} the intersection of the two boundaries while we will refer to this entire system as a \textit{wedge conformal field theory} (WCFT).

The WCFT setup naturally breaks the bulk full conformal symmetry $SO(d + 1, 1)$, in particular translations in the normal directions. 
The preserved amount of symmetry is then the conformal group in $(d-2)$ dimensions, namely $SO(d - 1, 1)$. Despite the conformal symmetry is significantly reduced from a global point of view, from an OPE perspective we can consider the system as containing four CFT's of different dimensionalities: one in the bulk with $SO(d + 1, 1)$ symmetry, one on each codimension-one boundary preserving $SO(d, 1)$, and one on the codimension-two wedge with $SO(d - 1, 1)$ symmetry.


 Quantum numbers of local operators in a WCFT are related to the surface and wedge critical exponents in addition to the bulk ones. Probing such CFT's not only constrains the surface and wedge observables, but also gives access to a part of the bulk observables in a homogeneous CFT (without boundaries). In particular, while the quantum numbers associated to the wall and the ramp are universal, the ones associated with the wedge depend on the angle $\th$. Systems with wedges have been studied in the older literature \cite{deutsch1979boundary, C1983, GT1984, barber1984magnetization, CR1984, kaiser1989surface, pleimling1998critical} and they are of interest to condensed matter physics. In addition, we also notice that the identification of the two boundaries, $\theta \rightarrow 2\,\pi$, leads to a theory in the presence of a conical singularity located on the wedge. Quantum field theories in the presence of such conical defects have been widely discussed in the literature, see e.g. \cite{dowker1987vacuum, de1989classical, alford1989aharonov, alford1989enhanced}. They play an important role in cosmology and string theory (cosmic strings) \cite{Kibble_1976, Copeland:2003bj}, and in the entanglement literature where they are used as a computational tool for studying Rényi and entanglement entropies  \cite{Callan:1994py, Solodukhin:1994yz}.

One powerful method to study and to characterise CFT's is the conformal bootstrap. This approach is used to constrain the space of possible conformal dimensions and OPE coefficients, by enforcing a few mild assumptions based mostly on symmetries. Since its revival in \cite{Rattazzi:2008pe}, there has been huge progress in using the conformal bootstrap in its numerical and analytic incarnation applied in several contexts, see \cite{Poland:2018epd,Bissi:2022mrs} for recent reviews on the topics. Recently, there has been an interest in studying WCFT's using the methods of conformal bootstrap \cite{Antunes:2021qpy}. This is a non-perturbative approach and refers to no Lagrangian description for the underlying microscopic theory.  The idea is to focus on the residual symmetries preserved by the wedge and the boundaries, and impose consistency conditions on the correlation functions. Although the symmetry is reduced from the full conformal group, it is still powerful enough to constrain the observables. This is a nice playground to study the bulk properties of the CFT in addition to the boundary and wedge ones. 
Bulk one-point functions are non-trivial observables in this case and admit an expansion into wall- and ramp-channel conformal blocks \cite{Antunes:2021qpy}. Moreover, two-point correlation functions of a bulk and an edge scalar operator carry information about the dynamical content of the CFT living on the wedge. Those correlators can also be decomposed into a wall- and a ramp-channel \cite{Antunes:2021qpy}, and the dynamical content of the WCFT data entering in these bootstrap equations is encoded in the bulk, boundary and wedge scaling dimensions in addition to the corresponding boundary operator product expansion (BOE) coefficients. 

Motivated by these developments, in this paper we focus on the study of a scalar field theory with global $O(N)$ symmetry in the free case first, and successively in the presence of either bulk or boundary interactions. Since we are interested in the situation with the maximal amount of symmetry, we will only impose conformal boundary conditions (b.c.'s) both on the wall and ramp. In particular, there are three distinct possible choices of b.c.'s that we can consider:
\begin{itemize}
	\item NN: Both boundaries are equipped with Neumann b.c.'s. 
	\item DD: Both boundaries are equipped with Dirichlet b.c.'s.
	\item DN: One boundary (for example the wall) is equipped with Dirichlet b.c., and the other (the ramp) with Neumann b.c.
\end{itemize}

 In this setup, we will study, among other things, the two-point functions involving one bulk and one wedge operator. More specifically, we compute the solution to the Dyson-Schwinger (DS) equation for a quartic interaction in the $d=4-\epsilon$ bulk and on one of the boundaries in $d=3-\epsilon$, up to order $\mathcal{O}(\epsilon)$, along the lines of  \cite{Liendo:2012hy}.  We have extracted the anomalous dimensions from such correlators, which we have complemented with Feynman diagrams computations, finding perfect agreement.  We have also used the OPE to extract BOE coefficients. At order $\mathcal{O}(\epsilon)$ the structure of the result is quite neat, and it contains a logarithmic term as expected at this order in perturbation theory. 

There are several directions that would be interesting to pursue. 
\begin{itemize}
\item \textit{Discontinuity as building block}: In homogeneous CFT's there has been progress in understanding four-point correlators using a single variable dispersion relation by exploiting the analytic structure of the bootstrap equation \cite{ Bissi:2019kkx, Bissi:2021spj}. The idea is to use the analytic structure together with the crossing symmetry of the correlator to constrain the CFT data. The correlators in a boundary conformal field theory (BCFT) can also be studied using bootstrap methods. The bootstrap approach for BCFT's was initiated in \cite{Liendo:2012hy}. In \cite{Bissi:2018mcq}, the analytic structure of the conformal blocks was exploited to constrain the BCFT data. This method was further modified in \cite{Dey:2020jlc} for CFT's with interfaces (where there is a bulk theory on each side of the codimension one defect). It would be interesting to generalise the approach presented recently in \cite{Bianchi:2022ppi,Barrat:2022psm} to the wedge configuration to be able to understand how to systematise the study of one- and two-point functions. As a consequence, it would be interesting to further explore the connections found in \cite{Sinha:2022sdo} to more general cases, such as the one presented in this paper.
\item \textit{Bootstrap equations}: Another interesting direction to pursue is to exploit the power of conformal symmetry and use the conformal bootstrap approach, for instance adapting \cite{Antunes:2021qpy} to higher order in perturbation theory. This approach would allow us to study more general interactions and to constrain different OPE data, using the different OPEs involving operators belonging to the wall/ramp/bulk. In addition, it would be interesting to complement the WCFT bootstrap equations with the ones coming from the BCFT and the homogeneous case. The study of the compatibility of the mixed system could potentially constrain even more the CFT data. 
\item \textit{Holographic description}: Holography in the form of the AdS/CFT correspondence \cite{Maldacena:1997re, Witten:1998qj, Gubser:1998bc} is one of the main tools to approach quantum field theories at strong coupling. The AdS space-time can be endowed with suitable branes which can be used to add boundaries and/or defects to the CFT living on the asymptotic boundary of the holographic space-time. In particular, a bottom-up construction has been proposed in \cite{Takayanagi:2011zk, Fujita:2011fp} where an end-of-the-world brane introduces a boundary in the CFT side of the correspondence. In this framework, it would be interesting to first find solutions to the Einstein equations describing CFT's confined in wedge-shaped regions,\footnote{In the context of the AdS$_3$/CFT$_2$ duality, the AdS-bulk dual to a wedge in $d=2$ has been found in \cite{Geng:2021iyq}.} and successively study their correlation functions and other relevant quantities related to quantum information.  The holographic approach is particularly convenient to obtain analytic and numerical results for the entanglement entropy by using the Ryu-Takayanagi formula \cite{Ryu:2006bv, Ryu:2006ef}, which can also be applied to the AdS/BCFT case \cite{FarajiAstaneh:2017hqv, Chu:2017aab,Seminara:2017hhh, Seminara:2018pmr}. The entanglement entropy has been computed in a free-scalar WCFT in \cite{Hertzberg:2010uv} (see also \cite{Berthiere:2018ouo} for a numerical approach). \\ \\
\item \textit{Conformal anomalies and RG flows}: Conformal anomalies and their associated coefficients, sometimes called central charges, play a major role in the classification of CFT's and in defining monotonic quantities along the RG flow, providing strong constraints on the possible outcomes of the latter. While in a CFT conformal anomalies are present only in even space dimensions, the situation is richer when boundaries and/or defects occur, and the possibility of having anomalies localised on them considerably extends the possible anomaly contributions \cite{Nozaki:2012qd, Solodukhin:2015eca, Fursaev:2015wpa, FarajiAstaneh:2021foi,Chalabi:2021jud}. Conformal anomalies of boundaries and defects have been widely studied in the literature, where monotonicity theorems have been proven \cite{Friedan:2003yc,Jensen:2015swa,Casini:2018nym, Cuomo:2021rkm,Wang:2021mdq} or conjectured \cite{Kobayashi:2018lil}, and various relations to boundary and bulk correlation functions have been found \cite{ Huang:2016rol, Herzog:2017xha, Herzog:2017kkj, Miao:2017aba, Jensen:2018rxu,Prochazka:2018bpb}. We find it interesting to investigate them in the context of the WCFT setup, which is a natural playground where anomalies localised on manifolds of different dimensions may relate to each other through either bulk, boundary or wedge RG flows.
\end{itemize}
 
The structure of the paper is as follows. 
In Sec. \ref{Sec:Free} we study a free scalar field theory confined in the wedge-shaped region described above in generic bulk dimension $d$. We first obtain the mode expansion of the field (\eqq{expansion_phi}) in the canonical formulation for all the possible combinations of b.c.'s, and by taking a suitable wedge limit we define some of the wedge primaries of the theory. After that, we find  the propagator reported in eq.'s (\ref{eq:propagator_NNDD}, \ref{eq:propagator_ND}), from which we obtain the relevant correlators we will use in the rest of the paper, namely the bulk one-point function of the fundamental field (\ref{eq:one-point_d4}, \ref{eq:one-point_d4_DN}, \ref{eq:one-point_gen_d}) and of the stress tensor (\ref{eq:stress_d_4_DDNN}, \ref{eq:stress_d_4_ND}, \ref{eq:stress_gen_d}). As regarding the latter, in \eqq{stress_general} we also report its general form consistent with conservation and tracelness. We then compute the bulk-wedge and wedge-wedge two-point functions that can be found in eq. \eqref{Bulk edge corr} and \eqref{eq:hhOhhO} respectively. We finally discuss a wedge-RG flow triggered by wedge primaries corresponding to a mode singular in the wedge limit which can be present in the expansion of the bulk fundamental field.

In Sec. \ref{Sec: EOM} we study the bulk-wedge two-point correlation function of scalar operators using the equation of motion (e.o.m.) for different b.c.'s. In this way, we are able to obtain the first-order correction of the bulk-wedge correlators  (eq.'s \eqref{Dyn fcn}) when the $\phi^4$ bulk-interaction is taken into account, and the anomalous dimension of the wedge primaries corresponding to either bulk- or boundary-interactions (see eq. \eqref{Wedge anom dim} and \eqref{3d wedge anom dim} respectively).

In Sec. \ref{review} we extract anomalous dimensions of wall and ramp fields by studying the respective boundary limits of the bulk-wedge correlator, and comparing it with the expected result found from the BOE. Moreover, as was showed in \cite{Antunes:2021qpy}, bulk-wedge correlators can be decomposed in wall- and ramp-conformal blocks. We find that these blocks satisfy the orthogonality relation \eqref{block orth}, which we use to read off the BOE coefficients entering in the bootstrap eq. (up to $\mco(\ep)$).

Part of the computations has been relegated to the appendices. In App. \ref{app:prop} we report the details of the derivation of the propagator in the free theory case. In App. \ref{app:feynman} we provide an alternative derivation of the wedge anomalous dimensions, and finally in App. \ref{App: Block} we give some detail on the expansion of conformal blocks at first order in $\epsilon$.

\section{Free scalar in a wedge} \label{Sec:Free}

In this Section we consider the theory of a single scalar confined in a wedge. We take the classical Euclidean action of the free scalar in $d$ dimension to be
\begin{equation}
S=  \frac{1}{2}  \int_\mathcal{M} d^d x \, \sqrt{g} \left(   g_{\mu\nu} \nabla^\mu \phi \nabla^\nu \phi + \frac{d-2}{4(d-1)} \mathcal{R}\,\phi^2 \right) \ ,
\end{equation}
where the non-minimal coupling to the metric is required by Weyl invariance.
From this action we obtain the e.o.m.
\begin{equation}
\label{eq:scalar_eqmotion}
\left(-\nabla^2 + \frac{d-2}{4(d-1)} \mathcal{R}\right)  \phi =0 \ ,
\end{equation}
and the stress tensor
\begin{equation}
\label{eq:stress-tensor_scalar}
\begin{split}
T_{\mu\nu} = &\nabla_\mu \phi \nabla_\nu \phi  -\frac{d-2}{4(d-1)} \left[ \nabla_\mu \nabla_\nu  +\frac{ g_{\mu\nu}}{d-2} \nabla^2 - R_{\mu\nu}  \right]\phi^2 \ .
\end{split}
\end{equation}
The goal of this section is to solve the theory living in a wedge for generic $d$. We will use canonical quantisation to obtain the mode expansion of the field, thus we will switch to Lorentzian signature. However, in the rest of the paper, we will use interchangeably the Lorentzian and Euclidean formulations, which are simply related by the replacement $t \rightarrow -i \tau$. 
 In order to take advantage of the symmetry of the problem, we find it convenient to adopt polar coordinates in the $(x_{d-1}, x_d)$ plane, namely 
\begin{equation}
ds^2 = -dt^2 + d \boldsymbol{x}^2 + d\rho^2 + \rho^2 d\varphi^2 \ , \quad x_d^2 + x_{d-1}^2 = \rho^2 \ , \quad \frac{x_{d-1}}{x_d} = \tan \varphi \ ,
\end{equation}
with $\rho > 0$ and $\varphi \in (0, \th)$. Furthermore we will denote the $d-2$ coordinates parallel to the boundaries as $x_\para = (t, \boldsymbol{x})$.

In this coordinate system the wedge is defined by $\mathcal{M}=\{ x_\para \in \R^{d - 2}, \, \rho > 0, \, \varphi \in (0,\theta)  \}$ where $\theta$ is the opening angle. We are interested in breaking the minimum amount of symmetries. In particular, this means we will impose only conformal b.c.'s. In the case of a free scalar, we can impose either Dirichlet ($\phi = 0$ at the boundaries) or Neumann ($\partial_\perp \phi = 0$ at the boundary) b.c.'s.

The e.o.m. in flat space-time reads
\begin{equation}
\begin{split}
\frac{1}{\rho} \partial_\rho (\rho\, \partial_\rho \phi) +\left[ -\partial_t^2 +\frac{\partial^2_\varphi}{\rho^2} + \partial_{\boldsymbol{x}}^2 \right]\phi =0\ ,
\end{split}
\end{equation}
which can be solved by the following ansatz (see for example \cite{deutsch1979boundary})
\begin{equation}
\phi  \sim e^{-i\, \omega\, t + i\,\boldsymbol{k} \cdot \boldsymbol{x}} h(\rho) g(\varphi) \ , \quad g_\alpha(\varphi) \equiv   A\, \sin (\alpha\, \varphi) + B\, \cos (\alpha\, \varphi) \ ,
\end{equation}
where the coefficients $A$ and  $B$ as well as the parameter $\alpha$ need to be fixed by imposing the b.c.'s (both on the wall and the ramp) and from the normalisation of the angular part $A^2+B^2=1$.

As discussed in Sec. \ref{sec:intro}, since we are free to impose different b.c.'s on the two boundaries we have 3 distinct possibilities: Dirichlet-Dirichlet (DD), Neumann-Neumann (NN), and Dirichlet-Neumann (DN). Imposing such conditions gives
\beq
\label{eq:g_a}
\begin{cases}
\displaystyle{A=1 \ , \quad B = 0 \ , \quad \alpha = \frac{m\, \pi}{\theta} \ ,} & \text{DD}, \\
\displaystyle{A=0 \ , \quad B=1 \ , \quad  \alpha = \frac{m\, \pi}{\theta} \ ,} &  \text{NN}, \\
\displaystyle{A= 1 \ , \quad B = 0 \ , \quad \alpha = \frac{\left(m  +\frac{1}{2} \right)\pi}{\theta} \ ,} & \text{DN}.
\end{cases}
\eeq
where $m$ is an integer (non-zero for DD).

This ansatz leads to an ordinary differential equation for $h(\rho)$, namely
\begin{equation}
\rho^2 h''+\rho \, h'+\left[(\omega^2-{\boldsymbol{k}}^2)\rho^2- \alpha^2 \right]h=0 \ ,
\end{equation}
whose solutions are the Bessel functions
\begin{equation}
h= J_{\pm \alpha} \left(\sqrt{\omega^2-{\boldsymbol{k}}^2} \rho \right) \ , \quad h= Y_{\pm \alpha} \left(\sqrt{\omega^2-{\boldsymbol{k}}^2} \rho\right) \ .
\end{equation}
Regularity of the solution at $\rho \rightarrow 0$ for any value of $\theta \in (0,2\pi)$ excludes $Y_{\pm\al}$, and imposes $\pm \alpha \ge 0$ for the Bessel $J_{\pm \alpha}$.\footnote{In Sec. \ref{Sec: RG}, we discuss about the possibility of having mildly divergent modes in the wedge limit.} This requirement leads to
\begin{equation}
\begin{aligned}
\label{eq:expansion_phi}
\phi(x) &=  \sum_{m \geq 0}{}^{\prime}  \int_{\R^{d - 2}} d^{d - 2}{\boldsymbol{k}}\int_0^{\infty} dk_\rho \, \,\frac{\sqrt{k_\rho}}{(\sqrt{2\,\pi})^{d-3}\sqrt{2\,\omega}}\sqrt{\frac{2}{\theta}} \bigg[  a_{\{k\}} \, g_{\alpha(m)}(\varphi) e^{-i\, \omega\, t +i\,  \boldsymbol{k} \cdot \boldsymbol{x}} + h.c. \bigg] J_{\alpha(m)} \left(k_\rho\,\rho\right) \ ,
\end{aligned}
\end{equation}
where we defined $\{k\} \equiv (\boldsymbol{k},k_\rho,m)$ and the primed sum means that the zero mode needs to be halved, i.e. its normalisation has an additional $1/\sqrt{2}$.
Note that in order to quantise the field, we found it convenient to change integration variable setting $k_\rho=\sqrt{\omega^2-{\boldsymbol{k}}^2}$. 
The  usual canonical commutation relation $[\phi,\Pi]=[\phi,\dot \phi]=i\,\delta$ implies the following commutation relations for the operators $a$ and $a^\dagger$\,\footnote{In the computation we use the following orthogonality property of the Bessel functions
\begin{equation}
\begin{aligned}
\int_{0}^\infty d\rho\, \rho J_\beta(\rho\, v)J_\beta(\rho\, u)=\frac{\delta(u-v)}{u}  \ .
\end{aligned}
\end{equation}}
\begin{equation}
\begin{aligned}
\label{eq:comm_rel}
\left[ a_{\{k\}}, a^\dagger_{\{k'\}}\right]=\delta({\boldsymbol{k}}-{\boldsymbol{k'}})\delta(k_\rho- k'_\rho)\delta_{m,m'}  \ .
\end{aligned}
\end{equation}
By applying the bulk-boundary BOE, followed by the boundary-wedge BOE, we can write the bulk field in terms of wedge modes \cite{Antunes:2021qpy}. For Dirichlet or Neumann b.c.'s there is only one boundary primary in the bulk-boundary BOE \cite{Liendo:2012hy}. This means that we can write the fundamental field \eqref{eq:expansion_phi} in terms of wedge primaries, which we denote $\hathat{O}_{\alpha}$, as\footnote{The differential operator generating the boundary descendants gives us $g_\al(\varphi)$.}
\begin{equation}
\begin{aligned}
\label{eq:phi_diff_op}
\phi(x) = \sum_{m \geq 0}{}^{\prime}  \rho^\alpha  B^{d - 1}_{\hathatexp{\De}_\al}\left( \rho^2 \partial^2_\para \right)   \hathat O_\alpha (x_\para)g_\alpha (\varphi) \ ,
\end{aligned}
\end{equation}
where we defined
\begin{equation}
\label{eq:defect_prim}
\hathat O_\alpha \equiv  \frac{1}{2^{\alpha}\G(\al + 1)} \sqrt{\frac{2}{\theta}} \int \frac{d^{d-1} k}{(\sqrt{2\,\pi})^{d-3}}  \,\frac{\sqrt{k_\rho}}{\sqrt{2\,\omega}} \, k_\rho^{\alpha} \left(  a_{\{k\}}  e^{-i\, \omega\, t +i\,  \boldsymbol{k} \cdot \boldsymbol{x}} + h.c. \right) \ .
\end{equation}
%
The differential operator is defined as\footnote{That this differential operator reconstructs the full bulk field $\phi(x)$ can be seen by observing that the Bessel functions can be expanded as
\begin{equation}
J_{\alpha}(x) = \left( \frac{x}{2} \right)^{\alpha} \sum_{j\geq 0} \frac{(-1)^j }{j! \, \Gamma \left(j+1+\alpha \right)}\left( \frac{x}{2} \right)^{2j} \ ,
\end{equation}
which implies
\begin{equation}
\frac{J_{\alpha}\left(x\right)}{x^{\alpha}} =\frac{1}{2^{\alpha} \Gamma\left( 1+ \alpha \right)} B^{d - 1}_{\hathatexp{\De}_\al}\left( \partial^2_x \right) e^{\pm i x} \ .
\end{equation}}
\begin{equation}
B^{d - 1}_{\hathatexp{\De}_\al} \left(\rho^2 \partial^2_\para \right) \equiv \sum_{j=0}^{+\infty} \frac{(-4)^{-j} (\rho^2 \partial^2_\para)^j}{j! (1 + \hathat\Delta_\alpha - \Delta_\phi)_j} \ ,
\end{equation}
where $(a)_j\equiv a(a+1)\dots (a+j-1)$ and $(a)_0 \equiv 1$ is the Pochhammer symbol.
The quantity $\hathat\Delta_\alpha$ is
\beq
\label{eq:Delta_a}
\hathat \Delta_\alpha = \De_\ph + \alpha \ , \quad \De_\ph = \frac{d - 2}{2} \ ,
\eeq
and it corresponds to the engineering conformal dimension of the operator  $\hathat O_\alpha$.

Thus, we see that at least in the specific case of the free theory we can reconstruct the bulk field $\phi$ in terms of wedge primaries. This is in agreement with the discussion in \cite{Lauria:2020emq} for a codimension-2 defect. However, we find it important to highlight that the simple form of \eqq{phi_diff_op} in the present discussion comes from the fact that the  linearity of the e.o.m. allows us to factorise the angular dependence even though the wedge breaks the rotational symmetry around the $\theta$-direction. A more general approach, which is discussed in \cite{Antunes:2021qpy}, is expanding the bulk operator in terms of boundary operators, and successively expand the latter ones in terms of operators living on the intersection between the boundaries.

It is straightforward to generalise the theory of a single real scalar to the one of $N$ scalars with $O(N)$ symmetry. This can be achieved by introducing an internal index $i\in \{1,\dots, N \}$ and requiring that the commutation relations \eqq{comm_rel} are diagonal. From now on, we will report the result corresponding to the more general $O(N)$ model.

\subsection{The bulk-bulk propagator}
The bulk-bulk propagator can be computed directly from the mode expansion \eqref{eq:expansion_phi}. The computation is analysed in detail in App. \ref{app:prop}. Here we just report the results. For DD and NN b.c.'s we find
\begin{equation}
\label{eq:propagator_NNDD}
\begin{split}
\left<\phi^i( x_\para,\varphi,\rho)\phi^j(0,\varphi',\rho')\right> =  \de^{ij} \frac{\pi}{\theta} \sum_{m\in\Z} G_S^{(|\frac{\pi\,m}{\theta}|)} (x_\para,\rho; x'_\para,\rho') \left( e^{i \frac{\pi}{\theta}m (\varphi- \varphi')} \pm e^{i \frac{\pi}{\theta}m (\varphi+ \varphi')}  \right) \ ,
\end{split}
\end{equation}
where $+$ corresponds to NN while $-$ to DD, and we remind the reader we have promoted the fields to be invariant under $O(N)$, with indices $i,j\in\{1,...,N\}$. The function $G_s^{(\nu)}$ is\footnote{We note that the functions in \eqq{G_S} correspond to the defect conformal blocks found in \cite{Gaiotto:2013nva} (see also \cite{Lauria:2020emq, Bianchi:2021snj}) in the case of monodromy defects. In particular, \eqq{propagator_NNDD} can be seen as an expansion in defect (or, more precisely, wedge) conformal blocks. However, we want to stress that in our case this precise form might be a consequence of the choice of quadratic boundary conditions, and we do not know whether this structure would persist in the presence of boundary interactions.}
%
\begin{equation}
\label{eq:G_S}
\begin{split}
G^{(\nu)}_S (x,x')
&=   \frac{\Gamma\left(\De_\ph + \nu\right)}{4\,\pi^{\frac{d}{2}} \Gamma\left(\n + 1\right)} \left(\frac{1}{\rho\, \rho'}\right)^{\De_\ph} \left(\frac{\xi}{2}\right)^{\De_\ph + \nu}  \phantom{}_2 F_1\left( \frac{\De_\ph + \nu}{2}, \frac{\De_\ph + \nu + 1}{2}; \nu + 1; \xi^2  \right) \ ,
\end{split}
\end{equation}
where $\De_\ph$ is the scaling dimension of the bulk scalar given by \eqref{eq:Delta_a}, and $\xi$ is the cross-ratio
\begin{equation}
\xi \equiv \frac{2\,\rho\,\rho'}{\rho^2 + \rho'^2+(x_\para - x_\para')^2} \ .
\end{equation}
In the DN case we find
\begin{equation}
\label{eq:propagator_ND}
\begin{split}
\left<\phi^i( x_\para,\varphi,\rho)\phi^j(0,\varphi',\rho')\right> =  \frac{4\,\pi\,\de^{ij}}{\theta}  \sum_{m\in\Z} G_S^{\left(\frac{\pi}{\theta} \left(m+\frac{1}{2}\right) \right)} (x_\para,\rho; x'_\para,\rho')\times & \\ 
\times \sin\left[\left( m + \frac{1}{2} \right) \frac{\pi}{\theta} \varphi  \right] \sin\left[\left( m + \frac{1}{2} \right) \frac{\pi}{\theta} \varphi'  \right]& \ .
\end{split}
\end{equation}
\subsubsection*{$\boldsymbol{d=4}$ case}
Interestingly, for $d=4$ the hypergeometric function appearing in \eqref{eq:G_S} simplifies to
\begin{equation}
\phantom{}_2 F_1\left( \frac{\nu+1}{2},\frac{\nu+2}{2}; \nu +1; \xi^2 \right) = 2^\nu \frac{\left( \sqrt{1-\xi^2} +1 \right)^{-\nu}}{\sqrt{1-\xi^2}} \ .
\end{equation}
This makes it possible to perform the sum over $m$ finding a result in closed form. For DD and NN we obtain
\begin{align} \label{eq:propNNDDd4}
\left<\phi^i(x) \phi^j(x') \right> = &\frac{\de^{ij}}{4\,\pi^2  }\frac{\pi}{\theta} \left(\frac{\xi}{2\,\rho\, \rho'}\right)  \sum_{m=0}^{+\infty}  \frac{ \chi^{-\frac{\pi}{\theta} |m|}}{\sqrt{1-\xi^2}} \left( e^{i \frac{\pi}{\theta}m (\varphi- \varphi')}  \pm e^{i \frac{\pi}{\theta}m (\varphi+ \varphi')}  \right) \\
= & \frac{\de^{ij}}{8\,\pi\, \theta } \frac{1}{\rho\, \rho'} \frac{\xi}{\sqrt{1-\xi ^2}}  \frac{\chi ^{\frac{2\, \pi }{\theta }}-1}{1+\chi ^{\frac{2\, \pi
   }{\theta }}-2\, \chi ^{\pi /\theta } \cos \left(\frac{\pi  (\varphi - \varphi')
   }{\theta }\right)} \pm \Bigg[(\varphi- \varphi') \rightarrow (\varphi + \varphi') \Bigg] \nonumber \ ,
\end{align}
where we defined
\beq
\chi \equiv \frac{\sqrt{1-\xi^2} +1}{\xi} \ ,
\eeq
while for DN b.c.'s we get
\begin{equation}
\label{eq:propNDd4}
\begin{split}
\left<\phi^i(x)\phi^j(x')\right> = \frac{\de^{ij}}{4\,\pi\, \theta } \frac{1}{\rho\, \rho'} \frac{\xi}{\sqrt{1-\xi ^2}} 
 \frac{ \chi^{\frac{\pi }{2\, \theta }} \left(\chi^{\frac{\pi}{\theta} }-1 \right) \cos \left( \frac{\pi (\varphi-\varphi')}{2\,\theta} \right)}{1+\chi ^{\frac{2\, \pi }{\theta }}-2 \chi ^{\frac{\pi}{\theta} }\cos \left( \frac{\pi (\varphi-\varphi')}{\theta} \right)} - \Bigg[ (\varphi- \varphi' ) \rightarrow  (\varphi + \varphi' )\Bigg] \ .
\end{split}
\end{equation}

\subsubsection*{$\boldsymbol{\theta=\frac{\pi}{n}}$ case in generic $\boldsymbol{d}$}

\begin{figure} 
	\centering
	\includegraphics[width=0.35\textwidth]{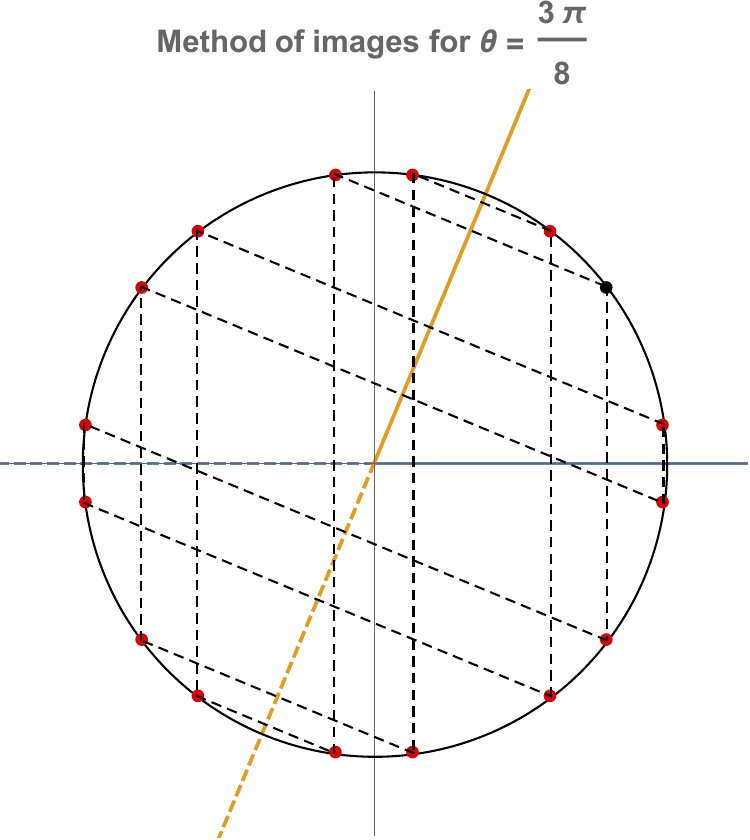}
	\caption{The $15$ image points (in red) required to solve the free theory DS equations for $\th = \frac{3\, \pi}{8}$. The blue line is the wall, and the orange is the ramp. The dashed lines clarify how the bulk point (in black) is reflected through the two boundaries. All of the image points lie on a circle.} 
	\label{Fig: im}
\end{figure}

Another case in which the sum over $m$ simplifies is when $\theta = \frac{\pi}{n}$, with $n \in \Z_{\geq 0}$ for DD and NN b.c.'s. Also in this case the details of the derivation are in App. \ref{app:prop}. The result is
\begin{align} \label{eq:phi-phi-images}
\left<\phi^i(x_\para,\varphi,\rho)\phi^j(0,\varphi',\rho')\right> &= \de^{ij}\frac{\Ga\left(\De_\ph\right)}{4\, \pi^\frac{d}{2}} \sum_{k = 0}^{n-1}  \left(\frac{1}{\left[ \rho^2 + \rho'^2 - 2\, \rho \,\rho' \cos \left(\frac{2\, k \,\pi}{n} + \varphi - \varphi' \right)+ x_\para^2 \right]^{\De_\ph}} \right. \nonumber \\
& \pm  \left. \frac{1}{\left[ \rho^2 + \rho'^2 - 2\, \rho\, \rho' \cos \left(\frac{2\, k\, \pi}{n} + \varphi + \varphi' \right) + x_\para^2\right]^{\De_\ph}}\right) \ .
\end{align}
Notice that this expression can also be derived directly by using the method of images \cite{sommerfeld1897vieldeutige,diehl1986field,Dowker:1977zj}. More generally, we can use method of images for angles $\th = \frac{a\,\pi}{b}$ with $a, b\in\Z_{\geq 0}$. Then if $\th \in (0, \pi]$ the number of image point required is $2\,\text{numerator}\left( \frac{b}{a} \right) - 1$.\footnote{In the case of $\pi < \th < 2\pi$, we need a finite $2\,\text{numerator}\left( \frac{b}{a - b} \right) - 1$ number of image points.} In Fig. \ref{Fig: im} we have an illustrative example.


\subsection{The one-point function of $\phi^2$}
From the bulk-bulk propagator we can compute the vacuum expectation value (v.e.v.) of $\phi^2$ in the case of $d=4$ for generic $\theta$, and in the case of $\theta = \frac{\pi}{n}$ for generic $d$. 
\subsubsection*{$\boldsymbol{d=4}$ case}

One way to proceed is by taking the coincident limit of the propagators (\ref{eq:propNNDDd4}, \ref{eq:propNDd4}).
In the case of DD and NN b.c.'s we find
\begin{equation}
\lim\limits_{y\rightarrow x}\left< \phi^i(x)\phi^j(y) \right> = \de^{ij}\left( \frac{1}{4\, \pi ^2 \varepsilon ^2} + \frac{1}{48\, \pi^2} \frac{\pi^2 -\theta^2 }{ \theta ^2 }\frac{1}{\rho^2} \pm \frac{\csc ^2\left(\frac{\pi  \,\varphi}{\theta }\right)}{16 \, \theta ^2 } \frac{1}{\rho ^2} + \mathcal{O}\left(\varepsilon^2\right) \right)\ ,
\end{equation}
where $\varepsilon$ is the regulator in the point-splitting procedure. The divergent piece proportional to $\varepsilon^{-2}$ represents the identity contribution in the $\phi$-$\phi$ bulk OPE that can be safely removed by subtracting the propagator in absence of boundaries. Thus, we obtain
\begin{equation}
\label{eq:one-point_d4}
\left< \phi^2(x) \right> =\frac{1}{48\, \pi^2} \frac{\pi^2 -\theta^2 }{ \theta ^2 }\frac{1}{\rho^2} \pm \frac{\csc ^2\left(\frac{\pi\,  \varphi}{\theta }\right)}{16 \, \theta ^2 } \frac{1}{\rho ^2}  \ .
\end{equation}
It is interesting to observe that the first term corresponds to the v.e.v. of the field in the presence of a conical singularity (see for example \cite{Bianchi:2015liz, Bianchi:2021snj}), while the second contribution is a modification due to the presence of the two boundaries.

The same procedure in the case of DN b.c.'s yields to 
\begin{equation}
\label{eq:one-point_d4_DN}
\left<\phi^2(x) \right> = -\frac{2 \,\theta ^2+\pi ^2 \left[1+6\, \cot \left(\frac{\pi\,  \varphi }{\theta }\right) \csc \left(\frac{\pi \, \varphi
	}{\theta }\right)\right]}{96\, \pi ^2 \theta ^2 \rho ^2} \ .
\end{equation}
The results for DD and NN have been found for the first time in  \cite{deutsch1979boundary}, while the result for DN is to the best of our knowledge novel.

Finally, as a limiting case we consider $\theta = \pi$, which gives
\begin{equation}
\left<\phi^2(x) \right>	=-\frac{2 \cot \varphi \csc \varphi +1}{32 \pi ^2 \rho ^2} \ , \quad \text{if } \theta = \pi \  .
\end{equation}
This correlator shows a non-trivial dependence on the coordinate $x_d$, which is parallel to the boundary. In addition, as a sanity check, we can see that in the near-boundary limit this one-point function behaves as the one with purely Neumann boundary condition if $x_d<0$ (or $\varphi>\frac{\pi}{2}$), and as the one with purely Dirichlet b.c.'s if $x_d>0$ (or $\varphi<\frac{\pi}{2}$) \cite{McAvity:1995zd}. This suggests the presence of an interface on top of the boundary at $x_d = 0$ (or $\varphi=\frac{\pi}{2}$) separating the two phases.

\subsubsection*{$\boldsymbol{\theta=\frac{\pi}{n}}$ case in generic $\boldsymbol{d}$}
In this case we obtain
\begin{equation}
\label{eq:one-point_gen_d}
\begin{split}
\left<\phi^2(x)\right> = \frac{\G({\De_\ph}) }{2^{\frac{d}{2}+1} \pi^\frac{d}{2}}  & \left\{\sum_{k = 1}^{n-1} \frac{1}{\left[ 1 -  \cos \left(\frac{2 \,k\, \pi}{n}  \right) \right]^{\De_\ph}}  \pm \sum_{k = 0}^{n-1}  \frac{1}{\left[ 1 -   \cos \left(\frac{2\, k\, \pi}{n} + 2\,\varphi \right) \right]^{\De_\ph}}\right\} \frac{1}{\rho^{2\,\De_\ph}} \ ,
\end{split}
\end{equation}
where $\De_\ph$ is given by \eqref{eq:Delta_a}. Unfortunately, we are not able to sum over $k$ in generic $d$. On the other hand, we can check the result \eqref{eq:one-point_d4} valid for $d=4$. 
The sums can then be analytically performed and we get
\begin{equation}
\left<\phi^2\right> = \frac{n^2-1 \pm 3\, n^2 \csc ^2(n\, \varphi  )}{48 \,\pi ^2} \frac{1}{\rho^2} \ , \quad d=4 \ ,
\end{equation}
which is exactly \eqref{eq:one-point_d4} for $\theta= \frac{\pi}{n}$.

\subsection{The one-point function of $T_{\mu \nu}$}

Another interesting correlator we can compute is the one-point function of the stress tensor. Conformal symmetry and conservation impose strict constraints to the possible form of this correlator. In particular, as we show below, we only need to find one particular component. 
In fact, by imposing scale invariance, rotations in the parallel directions of the wedge, and employing the fact that the stress tensor is symmetric we can write the one-point function as
\beq
\lan  T^{ij}\ran = \frac{f(\varphi)}{\rho^d} \delta^{ij} \ , \quad \lan T^{\rho\rho} \ran = \frac{g_{\rho\rho}(\varphi)}{\rho^d} \ , \quad \lan T^{\varphi\varphi} \ran = \frac{g_{\varphi\varphi}(\varphi)}{\rho^{d+2}} \ , \quad \lan T^{\varphi\rho} \ran = \frac{g_{\varphi\rho}(\varphi)}{\rho^{d+1}} \ .
\eeq
In addition, conservation leads to the following differential equations
\beq
\begin{split}
& g'_{\varphi \varphi}(\varphi )-(d-2) g_{\rho \varphi}(\varphi ) =0 \ , \\
& (1-d) g_{\rho \rho }(\varphi )- g_{\varphi \varphi }(\varphi
   )+ g'_{\rho \varphi}(\varphi ) = 0 \ ,
\end{split}
\eeq
while the condition of being traceless gives
\beq
(d-2) \, f(\varphi) + g_{\rho\rho}(\varphi) +  g_{\varphi\varphi}(\varphi) = 0 \ .
\eeq
Thus, we find that everything is fixed up to a single function $g(\varphi)$ as\footnote{We notice that the general form we obtained in \eqq{stress_general} is in conflict with the one reported in \cite{deutsch1979boundary}, which does not seem to be conserved to us. Regarding the specific case of the free scalar we find perfect matching.}
\beq
\label{eq:stress_general}
\begin{split}
& \left<  T^{ij}\right> = -\frac{1}{d-1} \left[ g(\varphi) + \frac{g''(\varphi)}{(d-2)^2} \right] \frac{1}{\rho^d} \delta^{ij} \ , \quad \left<T^{\rho\rho}\right> = -\frac{1}{d-1} \left[ g(\varphi) - \frac{g''(\varphi)}{d-2} \right]\frac{1}{\rho^d} \ , \\
&\quad \left<T^{\varphi\varphi} \right>= \frac{g(\varphi)}{\rho^{d+2}} \ , \quad \left<T^{\varphi\rho} \right>=\frac{1}{d-2} \frac{g'(\varphi)}{\rho^{d+1}}  \ .
\end{split}
\eeq
For this reason we find it convenient to compute the component $T_{\varphi\varphi}$. Its v.e.v. can be computed by using \eqq{stress-tensor_scalar} and taking the coincident limit of $\left<\phi \phi\right>$. 
As we did for $\phi^2$, we first compute the correlator when $d=4$, and then we consider the case in generic dimension but with $\theta = \frac{\pi}{n}$.
\subsubsection*{$\boldsymbol{d=4}$ case}
For DD and NN the one-point function reads
\beq
\label{eq:stress_d_4_DDNN}
\left< T_{\varphi\varphi} \right> =- \frac{1}{480\, \pi^2}\frac{\pi ^4-\theta ^4}{\theta ^4 }\frac{1}{\rho^2} \ ,
\eeq
while in the DN case we obtain
\beq
\label{eq:stress_d_4_ND}
\left< T_{\varphi\varphi} \right> = \frac{1}{3840\, \pi^2} \frac{8\, \theta ^4+7\, \pi ^4}{\theta ^4 } \frac{1}{\rho^2} \ .
\eeq
Note that in both the cases this correlator is independent of the angle $\varphi$. This means that also all the others components will be a function of $\rho$ only. Interestingly, we note that the sign of the coefficient of the stress tensor one-point function changes sign at $\theta = \pi$ only in the DD and NN cases, while in the DN case it remains negative for all the possible values of $\theta$. Also in this case, the results for DD and NN have been found for the first time in  \cite{Dowker_1978,deutsch1979boundary}, while the result for DN is to the best of our knowledge novel.\footnote{J.S. Dowker pointed out to us that, for free scalars, integrated quantities with DN b.c.'s can be obtained from the ones with DD and NN b.c.'s by applying simple algebraic identities \cite{Dowker:2004mq}. By applying that method to the correlator $\left< T_{\varphi\varphi} \right> \theta$ we find exactly \eqq{stress_d_4_ND}.}


\subsubsection*{$\boldsymbol{\theta=\frac{\pi}{n}}$ case in generic $\boldsymbol{d}$}
In this case, for the DD and NN case we find
\beq
\label{eq:stress_gen_d}
\left< T_{\varphi\varphi} \right>  =-\frac{\Gamma\left(\frac{d}{2}\right) }{2^{d+2}\pi^{\frac{d}{2}}} \frac{1}{\rho^{d-2}} \sum_{k=1}^{n-1} \frac{ (d-2) \cos \left(\frac{2\, \pi \,
   k}{n}\right)+d}{\sin^d\left(\frac{\pi\,  k}{n}\right)} \ .
\eeq

\subsection{Correlators of wedge operators}

From the general form of the propagators, we can compute the bulk-wedge correlators by employing the explicit form of the operators $\hathat O_\a$ reported in eq. \eqref{eq:defect_prim} (now promoted to $O(N)$). The computation is analogous to the one done for the full bulk-bulk propagator, thus we only give the result. We find
\begin{equation} \label{Bulk edge corr}
\begin{split}
\left<\phi^i(x) \hathat{O}^j_\a (x'_\para) \right> =& \frac{\Gamma\left(\hathatexp{\De}_\a\right)}{\pi^{\De_\ph} \theta \, \Gamma\left(\hathatexp{\De}_\a - \Delta_\phi + 1\right)} \frac{\de^{ij}}{\rho^{\De_\ph - \hathatexp{\De}_\a } (s_\para^2 + \rho^2)^{\hathatexp{\De}_\a} }\begin{cases}
\sin \left(\a \, \varphi\right)\ , &\quad \text{DD and DN}, \\
\cos \left( \a\, \varphi\right)\ , &\quad \text{NN},
\end{cases}
\end{split}
\end{equation}
where we remind the reader that $\Delta_\phi = (d - 2)/2$. Note also that even though the DD and DN cases seem identical they differ because of the possible values of $\a$, which are given in \eqq{g_a}.

Now we consider the wedge-wedge correlator
\begin{equation}
\label{eq:hhOhhO}
\left<\hathat{O}_\a^i \left(x_\para\right)\hathat{O}_{\a'}^j\left(x'_\para\right)\right> =  \frac{\Gamma\left(\hathatexp{\De}_\a\right)}{\pi^{\De_\ph} \theta \, \Gamma\left(\hathatexp{\De}_\a - \Delta_\phi + 1\right)} \frac{\delta^{ij}\delta_{\a\a'}}{\left| x_\para- x'_\para\right|^{2 \,\hathatexp{\De}_{\a}}} \ ,
\end{equation}
where $\hathat{\De}_{\a}$ is exactly the one introduced in \eqq{Delta_a}. 

\subsection{Singular modes and RG flows} \label{Sec: RG}
As anticipated it is possible to relax the requirement that $\phi$ is regular in the wedge limit $\rho \rightarrow 0$. In fact, this is allowed by unitarity in the following way: considering the wedge as a $(d-2)$-dimensional defect we can write the unitarity bound as \cite{Lauria:2020emq}
\begin{equation}
\begin{aligned}
\hathat \Delta &\ge \frac{d}{2} -2 \ , &\quad \text{if } d >4 \ , \\
\hathat \Delta &\ge 0\ , &\quad \text{if } d \le 4 \ .
\end{aligned}
\end{equation}
Note that the bound above is meant for operators different from the identity ($\hathat \Delta = 0$).
By employing the conformal dimensions \eqref{eq:Delta_a} we get
\begin{equation}
\begin{aligned}
\a &\ge -1\ ,  \quad &\text{if } &d>4\ , \\
\a &\ge 1 - \frac{d}{2}\ , \quad &\text{if } &d \le 4\ .
\end{aligned}
\end{equation}
Let us now focus for simplicity on the case $d=4$ with either DD or NN b.c.'s. In this case we see that there is a possible singular mode in the range $\theta \in (\pi, 2\,\pi)$ given by $\a^{(-)} = - \frac{\pi}{\theta}$.
Now we can trade the regular mode $\a^{(+)} = +\frac{\pi}{\theta}$ for $\a^{(-)}$. If we consider this new choice for the mode $|m|=1$, the spectrum of wedge operators will get modified in that we will have the operator $\hathat O_{-\frac{\pi}{\theta}}$ instead of $\hathat O_{+\frac{\pi}{\theta}}$. This, in turn, will modify the bulk-bulk propagator in the following way
\beq
\label{eq:prop_sub}
G_S^{(+\frac{\pi}{\theta})} \quad\Rightarrow\quad G_S^{(-\frac{\pi}{\theta})} \ .
\eeq
This modification also affects the one-point function of $\phi^2$
\beq
\left<\phi^2\right>^{(-)} - \left<\phi^2\right>^{(+)}  =  \frac{  \Gamma\left(\De_\ph+\frac{\pi }{\theta }\right) \Gamma \left(\De_\ph-\frac{\pi }{\theta }\right) \sin \left(\frac{\pi ^2}{\theta }\right)}{2^{d-3} \pi^{\frac{d+1}{2}} \Gamma\left(\frac{d-1}{2}\right)} \frac{1}{\rho^{d-2}}
\begin{cases}
\sin^2 \left( \frac{\pi}{\theta} \varphi \right) \ ,\quad & \text{DD,} \\
\cos^2 \left( \frac{\pi}{\theta} \varphi\right) \ ,\quad & \text{NN,}
\end{cases}
\eeq
while for the stress tensor we find
\beq
\left<T_{\varphi\varphi}\right>^{(-)} - \left<T_{\varphi\varphi}\right>^{(+)} = \frac{  (d-1)\sin \left(\frac{\pi
   ^2}{\theta }\right) \Gamma\left(\De_\ph+\frac{\pi }{\theta }\right) \Gamma\left(\De_\ph-\frac{\pi }{\theta }\right)}{ 2^{d-1}\pi ^{\frac{d - 5}{2}} \theta ^3 \, \Gamma\left(\frac{d+1}{2}\right)} \frac{1}{\rho^{d-2}} \ .
\eeq
Another interesting feature of this singular mode is that if we consider the quadratic relevant wedge perturbation
\beq
S \ni h \int d^{d-2} x_\para \, \hathat O_{-\frac{\pi}{\theta}} \hathat O_{-\frac{\pi}{\theta}} \ ,
\eeq
then we trigger an RG flow from the theory with the mode $\a^{(-)}$ to the one with the regular one $\a^{(+)}$. An analogous situation appears for a scalar in the presence of a monodromy defect as studied in \cite{Bianchi:2019sxz,Lauria:2020emq,Giombi:2021uae,Bianchi:2021snj}. In that case, unitarity allows for some modes which are mild divergent in the defect limit and provide different b.c.'s to the solution of the e.o.m for the field $\phi$. Different b.c.'s are shown to be related by defect RG flows, and in particular  in \cite{Bianchi:2021snj} the authors computed explicitly the bulk-bulk propagator along the defect RG flow. For the case of our interest, a very analogous computation can be performed, and  we obtain for $d=4$\footnote{We refer the reader to \cite{Bianchi:2021snj} for details on the computation.}
 \beq
 \begin{split}
\Delta\left<\phi^i(x_1) \phi^j(x_2)\right> &\equiv \left<\phi^i(x_1) \phi^j(x_2)\right>^{(+)} - \left<\phi^i(x_1) \phi^j(x_2)\right>^{(-)}  \\
&= - 4\,\de^{ij} \frac{\sin \left( \frac{\pi^2}{\theta} \right)}{\pi\, \theta} \int \frac{d^{2} k_\para}{(2\,\pi)^{2}} \frac{c \, h}{c \, h + k_\para^{\frac{2\,\pi}{\theta}}} K_{\frac{\pi}{\theta}} (k_\para \rho_1) K_{\pi/\theta} (k_\para \rho_2)\times \\
\eq \times e^{i\, k_\para(x_{||}^1-x_{||}^2)}
\begin{cases}
\displaystyle{\sin \left( \frac{\pi}{\theta} \varphi_1 \right) \sin \left( \frac{\pi}{\theta} \varphi_2 \right)} \ , & \text{DD}, \vspace{5px} \\
\displaystyle{\cos \left( \frac{\pi}{\theta} \varphi_1 \right) \cos \left( \frac{\pi}{\theta} \varphi_2 \right)} \ , & \text{NN},
\end{cases}
\end{split}
\eeq
where 
\beq
c \equiv \frac{4^{\frac{\pi}{\theta}} \Gamma \left( \frac{\pi}{\theta} \right)}{\theta \, \Gamma\left( 1 - \frac{\pi}{\theta} \right)} \ .
\eeq
We see that in the $h \rightarrow +\infty$ limit we obtain 
\beq
\label{eq:}
\begin{split}
\Delta\left<\phi^i(x_1) \phi^j(x_2)\right> &= - 4\,\delta^{ij} \frac{\sin \left( \frac{\pi^2}{\theta} \right)}{\pi\, \theta} \int \frac{d^{2} k_\para}{(2\,\pi)^{2}}  K_{\frac{\pi}{\theta}} (k_\para \rho_1) K_{\frac{\pi}{\theta}} (k_\para \rho_2) e^{i \,k_\para(x_{||}^1-x_{||}^2)} \times \\ 
\eq\times 
\begin{cases}
\displaystyle{\sin \left( \frac{\pi}{\theta} \varphi_1 \right) \sin \left( \frac{\pi}{\theta} \varphi_2 \right)} \ , & \text{DD}, \vspace{5px} \\
\displaystyle{\cos \left( \frac{\pi}{\theta} \varphi_1 \right) \cos \left( \frac{\pi}{\theta} \varphi_2 \right)} \ , & \text{NN}.
\end{cases}
\end{split}
\eeq
In this limit the integral can be performed analytically
\begin{equation}
\label{eq:diff_prop}
\Delta\left<\phi^i(x_1) \phi^j(x_2)\right> = - 4\,\de^{ij} \frac{\pi}{\theta} \left(\frac{1}{\rho_1 \rho_2} \right) \left[ \mathcal{F}\left(- \frac{\pi}{\theta} \right) - \mathcal{F}\left(\frac{\pi}{\theta} \right) \right] \begin{cases}
\sin \left( \frac{\pi}{\theta} \varphi_1 \right) \sin \left( \frac{\pi}{\theta} \varphi_2 \right) \ , & \text{DD}, \\
\cos \left( \frac{\pi}{\theta} \varphi_1 \right) \cos \left( \frac{\pi}{\theta} \varphi_2 \right) \ , & \text{NN},
\end{cases}
\end{equation}
where we defined
\beq
 \mathcal{F}(\a) \equiv \frac{1}{4\,\pi^{2}}  \left(\frac{\xi}{2}\right)^{1+ \a} {}_2F_1\left(  \frac{1+\a}{2}, 1+ \frac{\a}{2},\a ; \xi^2\right) \  .
\eeq
We notice that \eqq{diff_prop} produces exactly the substitution of \eqq{prop_sub}, proving that the two different b.c.'s are related by a wedge RG flow.

\section{The bulk-wedge correlator from the equation of motion} \label{Sec: EOM}

In this section, we study the bulk-wedge two-point functions of scalar operators, and how they can be found from the DS equations up to $\mco(\ep)$ with a quartic interaction in the bulk in $d = 4 - \ep$, or a quartic boundary interaction in $d=3-\ep$. This approach has been used  in \cite{Giombi:2020rmc, Giombi:2021cnr} in the context of a single boundary with a bulk interaction in $d = 4 - \ep$. 

In a WCFT there are four OPE's in play: two for each boundary, which allow us to express bulk operators in terms of the boundary operators, and two more
which express the boundary operators on the wall or the ramp in terms of wedge operators. We
call the former two the bulk-boundary BOE’s, and the latter two boundary-wedge BOE's. Using these OPE's we find that the bulk-wedge correlator for scalars is determined up to a function
\begin{equation}\label{bulkedge2ptfn}
\begin{aligned}
\langle \ph^i(x_\para, x_{d-1}, x_d) \hathat O^j(y_\para)\rangle = \frac{\de^{ij}f(\varphi)}{\rho^{\De_\phi - \Dhhe}\left(s_{||}^2 + \rho^2 \right)^{\Dhhe}} \, , \quad s_\para \equiv x_\para - y_\para \ .
\end{aligned}
\end{equation}
Here $\varphi$ is the single cross-ratio 
\begin{align} \label{crossratio2}
\varphi= \tan^{-1}\frac{ x_{d - 1}}{x_d} \ .
\end{align}
We will find $f(\varphi)$ by solving perturbatively the associated DS equations together with the different b.c.'s, which read\footnote{Note that the derivatives in $r$ in the directional derivatives w.r.t. the boundary normals vanish in the respective boundary limits.}
\begin{equation}
\begin{aligned}\label{bcall}
&{\text{DD:}} \quad f^{(0)}(0) = f^{(0)}( \theta)=0\ , \\
&{\text{NN:}} \quad \partial_{{\varphi }}f^{(0)}(0) = \partial_{\varphi }f^{(0)}( \theta)=0 \ , \\
&{\text{DN:}} \quad f^{(0)}(0) = \partial_{{\varphi }} f^{(0)}(\theta) =0 \ .
\end{aligned}
\end{equation}
The strategy we will follow is to expand all the relevant quantities such as the function $f$ and the scaling dimensions in terms of the dimensionless parameter $\epsilon$. Up to first order we write
\begin{equation} \label{eps exp}
\begin{aligned}
f(\varphi)&=f^{(0)}(\varphi)+\ep\, f^{(1)}(\varphi )+ \mco(\ep^2) \ , \\
\Delta _{\phi }&= \Delta _{\phi }^{(0)}+ \ep\,\g_{\ph} + \mco(\ep^2) \ , \\ 
\Dhhe&=\Dhhe^{(0)}+ \ep\,\hhg+ \mco(\ep^2) \ .
\end{aligned}
\end{equation}
Then, we need to plug the expansion \eqref{eps exp} into the DS equation for the field bulk field $\phi(x)$ and solve the differential equation order by order in $\epsilon$. Finally, the last step is to suitably impose the b.c.'s \eqref{bcall}, as explained in detail later.

\subsection{Free theory} \label{Sec: Free EOM}

To start we analyse the free theory case.
A massless scalar satisfies the Klein-Gordon equation
\begin{equation}
\begin{aligned}
\Box_x\ph^i = 0 \ ,
\end{aligned}
\end{equation}
which gives us the DS equation for the bulk-wedge correlator
\begin{align}\label{freeeom}
\Box_x \langle \ph^i(x_\para, x_{d-1}, x_d) \hathat O^j(y_\para)\rangle =0 \ .
\end{align}
When applied to \eqref{bulkedge2ptfn} we find the following differential equation 
\begin{align}\label{boxfree}
\de^{ij}\frac{\rho^{- \Delta _{\phi }^{(0)} +\Dhhe^{(0)}-2}}{\left(s_{||}^2 + \rho^2\right){}^{ \Dhhe^{(0)}}}
\left[ \pa_\varphi^2f^{(0)}(\varphi  )+f^{(0)}(\varphi  )\left(\left(\Dhhe^{(0)}- \Delta _{\phi }^{(0)}\right)^2-\frac{2 \rho^2 \,\Dhhe^{(0)} \left(d-2-2 \Delta _{\phi }^{(0)}\right)}{\rho^2+s_{||}^2 }\right)\right]=0 \ .
\end{align}
Since, as discussed above, conformal symmetry implies that $f$ is a function of $\varphi$ only, the second term inside the bracket of \eqref{boxfree} needs to vanish, which in turn implies 
\begin{align}\label{bulkdim0}
\Delta^{(0)} _{\phi }=\frac{d-2}{2} \ .
\end{align}
We notice that this is indeed the scaling dimension of the fundamental free bulk scalar field in \eqref{eq:Delta_a}. At this point we are left with the following differential equation for $f^{(0)}(\varphi  )$
\begin{align}
&\pa_\varphi^2f^{(0)}(\varphi  )+\left(\Dhhe^{(0)}- \Delta _{\phi }^{(0)}\right)^2\,f^{(0)}(\varphi  )=0 \ ,
\end{align}
with the general solution
\begin{equation} \label{Free gen sol}
\begin{aligned}
f^{(0)}(\varphi) &= A\cos \left[\left( \Dhhe^{(0)} - \De_\ph \right)\varphi\right] + B \sin \left[\left(\Dhhe^{(0)} - \De_\ph\right)\varphi \right] \ .
\end{aligned}
\end{equation}
Imposing the wall b.c. fixes one of the constants $A$  or $B$ to zero. The ramp b.c. will in turn give us the wedge scaling dimension
\begin{align}\label{edgescalingdim0}
\Dhhe^{(0)}_\al = \frac{d-2}{2}+\al\ , \quad \al= 
\frac{\pi }{\theta}\begin{cases}
\displaystyle{m} \ ,     \quad  & \text{DD, NN,} \\
\displaystyle{m+\half}\ ,\quad  & \text{ND,}
\end{cases}
\end{align}
with $m \in \Z$ (non-zero for DD). This agrees with \eqref{eq:Delta_a}. The remaining constant can be fixed by normalization, which we choose in such a way that we find agreement with \eqref{Bulk edge corr}
\begin{align} \label{Free dyn fcn}
f^{(0)}(\varphi )= \frac{\G({\hathatexp{\De}_\al})}{\pi^{\De_\ph}\th\,\G({\al + 1})}
\begin{cases}
\sin \left(\al\, \varphi  \right) \ ,\quad &{\text {DD, DN,}}\\
\cos \left(\al\, \varphi \right) \ ,\quad &{\text {NN.}}
\end{cases}
\end{align}
So far we have discussed the free scalar field in generic dimensions. Now we move on to discuss interacting Wilson-Fisher theories.

\subsection{Interacting theory: $\ph^4$-deformation in the bulk in $d = 4 - \ep$} \label{Sec: bulk int}

In $d = 4 - \ep$ we consider a quartic interaction in the bulk
\begin{equation}
\begin{aligned}
S \ni \int_{\R^d}d^dx\frac{\la}{8}(\ph^2)^2 \ , 
\end{aligned}
\end{equation}
from which we have the e.o.m.
\begin{align} \label{eom}
\Box_x\ph^i = \frac{\la}{2}\ph^2\ph^i  \ ,
\end{align}
where we denoted $\ph^2 \equiv (\ph^i)^2$. In the conformal case we have to specify the coupling at the RG fixed point
\begin{equation}
\begin{aligned}
\lambda_{*} = \frac{(4\,\pi)^2}{N + 8}\ep+\mco(\ep^2) \ . 
\end{aligned}
\end{equation}
The e.o.m. applied to the bulk-edge correlator \eqref{Bulk edge corr} is now given by
\begin{align}\label{eomepsilon}
\Box_x \langle \ph^i(x_\para, x_{d-1}, x_d) \hathat O_\al^j(y_\para)\rangle =\frac{\lambda_{*}}{2}  \langle \ph^2\ph^i(x_\para, x_{d-1}, x_d)\hhO_\al^j(y_\para)\rangle \ .
\end{align}
Since the coupling starts at $\mco(\ep)$, it is sufficient to consider the free theory correlators on the RHS. Applying Wick's theorem
\begin{align}\label{eomepsilon2}
\Box_x \langle \ph^i(x_\para, x_{d-1}, x_d) \hathat O^j_\al(y_\para)\rangle_1 = \frac{(N + 2)\lambda_{*}}{2} \langle \ph^i(x_\para, x_{d-1}, x_d) \hathat O^j_\al(y_\para)\rangle_{0}  \langle \ph^2(x_\para, x_{d-1}, x_d)\rangle_{0} \ ,
\end{align}
where we use the subscript $0$ to denote the free theory correlator, and $1$ is its first-order correction in the interacting theory. E.g. the differential equation at $\mco(\ep)$ for DD is given by
\begin{align}\label{phiconstraint}
&\pa_\varphi^2f^{(1)}+\al^2\, f^{(1)}(\varphi)+\g_\f  \frac{4\,x_d^2\,(\al+1)\sin \left(\al\, \varphi  \right)}{\pi\, \theta(x_d^2+s_{||}^2\cos ^2\varphi)}+\left(\theta ^2+3\, \pi ^2 \csc ^2\frac{\pi\,  \varphi }{\theta }-\pi ^2\right)\frac{ \sin \left(\al\, \varphi  \right) }{3\, \pi\,  \theta ^3}\nn
&+\frac{2\, \a (\g_\f-\hhg)}{\pi\, \theta}  \frac{\csc\varphi}{\tan \varphi } \sin \big((\al-1) \varphi \big)+\frac{(2\,\al-1)(\g_\f-\hhg)}{\pi\, \theta}\frac{\sin \left(\al\, \varphi  \right)}{\sin ^2\varphi }  =0
\end{align}
For it to only depend on $\varphi$ we have
\begin{align}\label{phi0}
\g_{\ph}=0 \ .
\end{align}
Similarly, from the differential equation at  $O(\ep)$  for NN and DN we find that this holds. Using this condition,
the differential equation at $O(\ep)$ reads 
\begin{align} \label{Diff eq}
 &\pa_\varphi^2f^{(1)}(\varphi )+\al^2 f^{(1)}(\varphi )+h(\varphi)=0 \ ,
\end{align}
where
\begin{align*}
h(\varphi)=
\begin{cases}
\displaystyle{\sin(\al\, \varphi) \left( \frac{2\,\pi\,\al}{\theta}\,\hhg
 +\left(\theta ^2+3\, \pi ^2 \csc ^2\frac{\pi\,  \varphi }{\theta }-\pi ^2\right) \frac{N + 2}{6(N + 8)\pi\,\theta^3} \right)} \ , \quad &{\text{DD,}} \\
\displaystyle{\cos  \left(\al\, \varphi  \right) \left( \frac{2\,\pi\,\al}{\theta}\,\hhg  +\left(\theta ^2-3\, \pi ^2 \csc ^2\frac{\pi\,  \varphi }{\theta }-\pi ^2\right) \frac{N + 2}{6(N + 8)\pi\,\theta^3} \right)} \ , \quad &{\text {NN,}} \\ 
\displaystyle{\sin  \left(\al\, \varphi  \right) \left( \frac{2\,\pi\,\al}{\theta}\,\hhg
 +\left(\theta ^2+3\, \pi ^2 \csc \frac{\pi  \,\varphi }{\theta }\cot \frac{\pi\,  \varphi }{\theta }+\frac{\pi ^2}{2}\right) \frac{N + 2}{6(N + 8)\pi\,\theta^3} \right)} \ , \quad &{\text {DN.}}
\end{cases}
\end{align*}
Note in particular that the $f^{(1)}(\varphi )$-term vanishes in eq. \eqref{Diff eq} for NN when $\al = 0$. This means that this solution has to be studied on its own, which we will do in the next subsection. 

For $\al \ne 0$, these equations can be solved for different values of $\al$ as given in \eqref{edgescalingdim0}. Similar to the free theory, the solution is fixed up to two constants. One of these constants is zero which is seen by implementing the wall b.c.\footnote{For Neumann b.c.'s there is a single pole in $\vph$, which corresponds to the BOE exchange of $\hp$. This term does not need to be zero. The b.c. is implemented by setting the $\vph^0$-term to zero. Similarly for NN, we encounter a single pole in $\th - \vph$ in the ramp limit of $\pa_\vph\ph$ which should not be tuned to zero.} By implementing the ramp b.c. we fix the edge anomalous dimension to be\footnote{We had to solve DN for $m = \frac{1}{2}$ and $m\geq \frac{3}{2}$ (in $\al$ of \eqref{edgescalingdim0}) as two separate cases.}
\begin{equation} \label{Wedge anom dim}
\begin{aligned}
\hhg=
\begin{cases}
\displaystyle{\frac{(N + 2)(\pi^2 - 6\,\pi\,\al\,\theta - \theta^2)}{12(N + 8)\al\,\theta^2}} \ , \quad &\text {DD, NN,} \\
\displaystyle{-\frac{(N + 2)(\pi^2 + 12\,\pi\,\al\,\theta + 2\,\theta^2)}{24(N + 8)\al\,\theta^2}} \ , \quad &\text {DN.} \\
\end{cases}
\end{aligned}
\end{equation}
The anomalous dimension for DD with $\alpha = \frac{\pi}{\theta}$ agrees with the older literature \cite{C1983}. We double check all of these anomalous dimensions using Feynman diagrams in App. \ref{sec:feynm_bulk}.
The final results are
\begin{equation} \label{Dyn fcn}
\begin{aligned}
f^{(1)DD}_{m\geq 1}(\varphi )&= - \pi^2 \left( \frac{N + 2}{2(N + 8) \pi\,  \theta } \log \left[ \sin \left(\frac{\pi\,  \varphi }{\theta } \right) \right] - \mathcal{A} \right) \sin(\al\,\vph) + \\ 
\eq + \pi \frac{N + 2}{2(N + 8) m\,  \theta}\sum_{i=1}^{m}\left(\frac{m}{i}-1\right)\sin \left(\frac{\pi(2\, i-m) \varphi  }{\theta } \right) \ ,
\end{aligned}
\end{equation}
\vspace{1px}
\begin{equation} 
\begin{aligned}
f^{(1)NN}_{m\geq 1}(\varphi )&= - \pi^2 \left( \frac{N + 2}{2(N + 8)\pi\,\th} \log \left[ \sin \left(\frac{\pi \, \varphi }{\theta } \right) \right] - \mathcal{A} \right) \cos(\al\,\vph) + \\
\eq - \pi^2 \frac{N + 2}{2(N + 8) \al\,\th^2}\sum_{i=1}^{m}\left(\frac{m}{i}-1\right)\cos \left(\frac{\pi(2\, i-m)\vph}{\th} \right) \\
\eq - \pi^2 \frac{N + 2}{2(N + 8)\al\,\th^2} \csc \left(\frac{\pi \, \varphi }{\theta }\right) \sin \left( \frac{\pi(m - 1)\varphi }{\theta } \right) \ ,
\end{aligned}
\end{equation}
\vspace{1px}
\begin{equation}
\begin{aligned}
f^{(1)ND}_{m = 0}(\vph) &= -\pi^2 \left( \frac{N + 2}{2(N + 8) \pi  \theta } \log \left[ \sin \left(\frac{\pi  \varphi }{\theta } \right) \right] - \mathcal{A} \right) \sin(\al\vph) \ ,
\end{aligned}
\end{equation}
\vspace{1px}
\begin{equation} 
\begin{aligned}
f^{(1)ND}_{m\geq 1}(\vph) &= -\pi^2 \left( \frac{N + 2}{2(N + 8) \pi \, \theta } \log \left[ \sin \left(\frac{\pi \, \varphi }{\theta } \right) \right] - \mathcal{A} \right) \sin(\al\,\vph) +                                                                                                                                       \\
\eq +\pi^2 \frac{N + 2}{4(N + 8)\al\,\th^2} \sum_{i=1}^{m}\left(\frac{2\, i-2\, m-1}{i}\right)\sin \left[ \left(m-2\, i+\half\right)\frac{\pi \, \varphi }{ \theta }\right] \\
\eq - \pi^2 \frac{N + 2}{4(N + 8)\al\,\th^2} \tan \left(\frac{\pi\,  \varphi }{2\, \theta }\right)\cos \left[\left(m+\frac{1}{2}\right)\frac{\pi  \, \varphi }{\theta }\right] \ .
\end{aligned}
\end{equation}
$\mathcal{A}$ is an undetermined constant which we will relate to a BOE coefficient in Sec. \ref{review}.

\subsubsection{NN $\al = 0$}

We solve the differential equations for the NN case with $\alpha=0$ 
\begin{align}
\frac{\partial^2}{\partial^2\varphi}f^{(0)}(\varphi)&=0\,,\nonumber\\
\frac{\partial^2}{\partial^2\varphi}f^{(1)}(\varphi)+\hat{\hat{\gamma} }^2_0 f^{(0)}(\varphi)+\frac{\theta^2-\pi^2-3\pi^2\csc^2(\frac{\pi \varphi}{\theta})}{18\pi \theta^3}&=0\,,
\end{align}
The solution at $O(0)$ reads
\begin{align}
f^{(0)}(\varphi)&= A+ B \varphi\,.
\end{align}
Using Neumann b.c. at $\varphi=0, \theta$ we obtain
\begin{align}
f^{(0)}(\varphi)&=A\,.
\end{align}
Using the normalisation we get 
\begin{align}
    A=\frac{1}{\pi \theta}\,.
\end{align}
The solution at $O(\epsilon)$ is then given by
\begin{align}
f^{(1)}(\varphi) &=C+ D \varphi+\frac{1}{36\pi \theta^3}\bigg(\left( \pi^2-\theta^2- 18 A \pi \theta^3 \hat{\hat{\gamma} }^2_0\right)\varphi^2-6\theta^2\log\sin\frac{\pi \varphi}{\theta}\bigg)\,.
\end{align}
The normal derivatives at $\varphi=0, \theta$ is 
\begin{align}
\partial_{\varphi}f^{(1)}(\varphi) &=D-\frac{1}{6\pi \theta \varphi}\,,\nonumber\\
\partial_{\varphi}f^{(1)}(\varphi) &=D-\frac{1}{6\pi \theta (\varphi-\theta)}+\frac{\pi^2-\theta^2- 18A \pi \theta^3 \hat{\hat{\gamma} }^2_0}{18 \pi \theta^2}\,.
\end{align}
Imposing the $O(\varphi^0)$ and $O((\varphi-\theta)^0)$ terms to zero we obtain
\begin{align}
D&=0\,,\nonumber\\
\hat{\hat{\gamma} }^2_0&=\frac{\pi^2-\theta^2}{18 \theta^2}\,.
\end{align}
This anomalous dimension is in agreement with \cite{Diatlyk:2024zkk}.\footnote{The paper \cite{Diatlyk:2024zkk} pointed out a disagreement with a previous version of this paper, that we fixed in the current one.}

\subsection{Interacting theory: $\hp^4$-deformation on the boundary in $d = 3 - \ep$}

In $d = 3 - \ep$ we can consider a quartic interaction on the boundary \cite{eisenriegler1988surface}, see also \cite{Prochazka:2019fah}. In our case we will consider it on the ramp
\begin{equation}
\begin{aligned}
S \ni \int_{\R^{d - 1}}d^{d - 1}x_\para\frac{g}{8}(\hp^2)^2 \ , 
\end{aligned}
\end{equation}
which gives us a modified Neumann b.c.'s on the ramp\footnote{As seen by varying the fields.}
\begin{equation}
\begin{aligned}
\frac{1}{\rho}\pa_\varphi \phi^i\Big|_{\vph = \theta} = -\frac{g}{2}\hp^2\hp^i \ .
\end{aligned}
\end{equation}
Here we denoted $\hp^2 \equiv (\hp^i)^2$. In the conformal case we have to specify the coupling to the RG fixed point
\begin{equation}
\begin{aligned}
g_* = \frac{4\,\pi}{N + 8}\ep + \mco(\ep^2) \ .
\end{aligned}
\end{equation}
Also in this case we will focus on the correlator \eqref{bulkedge2ptfn}, where we are interested in the first order in the $\epsilon$-expansion.
Since the bulk DS equation is not affected by the boundary perturbation, the solution to the function $f(\varphi)$ is the same as we found in eq. \eqref{Free gen sol}. The only difference is in the b.c. on the ramp, which now reads
%
\begin{equation}
\label{eq:bc_ramp_jnter}
\begin{aligned}
\frac{1}{\rho}\pa_\vph\langle \ph^i(x_\para, \rho, \varphi) \hathat O_\a^j(0)\rangle\Big|_{\vph = \theta} = -\delta^{ij} \frac{(N + 2)g_{*}}{2} \langle \hat \ph^i(x_\para, \rho) \hathat O^j_\al(0)\rangle_{0}  \langle \hat\ph^{i \,2}(x_\para, \rho)\rangle_{0} \ .
\end{aligned}
\end{equation}
As clear from the equation above, we need the boundary-wedge correlator and the one-point function of $\hat \phi^2$ in the free limit. While the former can be found in any dimension from the bulk-wedge correlator \eqref{Bulk edge corr}, we are able to report the form of the coefficient of the one-point function for $d=3$ only in the special case $\theta = \frac{\pi}{n}$. In particular, we have
\beq
\label{eq:bound_wedge}
\left<\hat\phi^i(x) \hathat O^j_\al (y_{||}) \right> =\delta^{ij}\frac{\Gamma\left(\alpha +\frac{1}{2}\right)}{\pi^{\frac{3}{2}} \theta\, \Gamma\left(\alpha +1\right)}   \frac{\rho^\a}{ \left( \rho^2 + (s_{||})^2\right)^{\a + \frac{1}{2}}} \ ,
\eeq
and
\begin{equation}
\label{eq:bdy_phi_sq}
\begin{split}
\left<\hat\phi^{2}\right> = \frac{1}{2 \sqrt{2}\, \pi }   \frac{\k_n}{\rho}, \qquad \k_n \equiv \sum_{k = 1}^{n-1} \frac{1}{\sqrt{ 1 -  \cos \left(\frac{2\, k\, \pi}{n}  \right)}} \ .
\end{split}
\end{equation}
By imposing the purely Neumann b.c. on the wall boundary fixes the coefficient $B = 0$. As in the free case, the other coefficient ($A$) is related to the normalisation of the wedge operators, and the b.c. on the ramp in \eqq{bc_ramp_jnter} fixes the conformal dimension of the wedge operator. By plugging the correlators in \eqref{eq:bound_wedge} and \eqref{eq:bdy_phi_sq} back to \eqq{bc_ramp_jnter}, and setting $d=3-\epsilon$, we obtain the anomalous dimension
\beq \label{3d wedge anom dim}
\hathat\gamma =  \frac{N+2}{N+8}\frac{ 1}{\sqrt{2}\, \alpha\, \theta  } \kappa_n \ .
\eeq
To check this result, we found it worth deriving it also from the  standard diagrammatic approach in App. \ref{app:wedge_anom_d3}.  

\section{CFT data at order $\ep$}\label{review}

In this section, we will explain how the BOE coefficients at $\mco(\ep)$ can be found from the bulk-wedge correlators found in Sec. \ref{Sec: bulk int}. This method relies on the conformal block decomposition from \cite{Antunes:2021qpy}, and in the process we will also find the anomalous dimension of the single boundary operator from the free theory ($\hp$ or $\pa_\perp\hp$).

\subsection{Boundary anomalous dimensions}

Using the bulk-boundary and boundary-wedge BOE's (for the wall) we can write the bulk-wedge correlator as
\begin{equation} \label{Bulk-wedge corr}
\begin{aligned}
\langle\ph^i(x)\hhO^j(y_\para)\rangle &= \sum_{\hO}\frac{\m^\ph{}_{\hO}}{|x_{d - 1}|^{\D_\ph - \hD}}B^d_{\hD}(x_{d - 1}^2)\langle\hO^i(x_\para, r)\hhO^j(y_\para)\rangle \ .
\end{aligned}
\end{equation}
The sum runs over all boundary primaries. The boundary-wedge correlator is fixed by conformal symmetry (in the same way as a bulk-boundary correlator in a BCFT)
\begin{equation}
\begin{aligned}
\langle\hO^i(x_\para, r)\hhO^j(y_\para)\rangle = \de^{ij}\frac{\hat{\m}^{\hO}{}_{\hhO}}{\rh^{\hD - \Dhhe}(s_\para^2 + \rh^2)^{\Dhhe}} \ .
\end{aligned}
\end{equation}
$B^d_{\hD}$ is a differential operator generating towers of descendants for each boundary primary
\begin{equation}
\begin{aligned}
B^d_{\hD}(x_{d - 1}^2) &= \sum_{m\geq 0}\frac{(-1)^mx_{d - 1}^{2m}}{m!\left( \hD - \frac{d - 3}{2} \right)_m}(\pa_{x_\para}^2 + \pa_{x_d}^2)^m \ .
\end{aligned}
\end{equation}
If we perform the summation over $m$ we find the conformal blocks \cite{Antunes:2021qpy}. However, for the purpose of finding the boundary anomalous dimensions from \eqref{Dyn fcn} it is enough to study eq. \eqref{Bulk-wedge corr} at the lowest order in $\rh$
\begin{equation} 
\begin{aligned}
\langle\ph^i(x)\hhO^j(y_\para)\rangle &= \de^{ij}\frac{c_{\hD}\sin^{\hD - \De_\ph}\vph}{\rh^{\D_\ph - \Dhhe}(s_\para^2 + r^2)^{\Dhhe}} + ... \ , \quad c_{\hD} = \m^\ph{}_{\hO_f}\hat{\m}^{\hO_f}{}_{\hhO} \ ,
\end{aligned}
\end{equation}
where $\hO_f \in \{\hp, \pa_\perp\hp\}$ is the boundary operator from the free theory. By comparing this with \eqref{bulkedge2ptfn} we find
\begin{equation} 
\begin{aligned}
f(\vph) &= c_{\hD}\sin^{\hD - \De_\ph}\vph + ... \ .
\end{aligned}
\end{equation}
In the $\ep$-expansion \eqref{eps exp} with\footnote{The boundary anomalous dimensions correspond to normal derivatives. This choice of scaling dimensions is motivated by the generalized free theory solutions \cite{Liendo:2012hy}.}
\begin{equation} \label{eps exp 2}
\begin{aligned}
\hD_n &= \D_\ph^{(0)} + n + \ep\,\hg_n + \mco(\ep^2) \ , \\
c_{\hD_n} &= c_{\hD_n}^{(0)} + \ep\, c_{\hD_n}^{(1)} + \mco(\ep^2) \ ,
\end{aligned}
\end{equation}
we have to the lowest order in $\vph$
\begin{equation} 
\begin{aligned}
f^{(1)}(\vph) &= c_{\hD_n}^{(0)}\vph^{\hD^{(0)}_n - \De_\ph^{(0)}} + ... \ , \\
f^{(1)}(\vph) &= \vph^{\hD^{(0)}_n - \De_\ph^{(0)}} ( c_{\hD_n}^{(1)} + c_{\hD_n}^{(0)}(\hg_n - \g_\ph)\log\vph ) + ... \ .
\end{aligned}
\end{equation}
If we were to use the BOE's for the ramp in \eqref{Bulk-wedge corr} we simply exchange $\vph \rightarrow \th - \vph$. 

By expanding our results of $f(\vph)$ from Sec. \ref{Sec: EOM} in $\vph$ (either around zero or $\th$) we can extract the CFT data of the boundary operator from the free theory. For  the three different combinations of b.c.'s we find the boundary anomalous dimensions \eqref{Bdy anom dim} (it is the same on the wall and the ramp) consistent with the BCFT literature. Remarkably, we have now found all of the anomalous dimensions appearing at $\mco(\ep)$ (bulk, boundary and wedge) only assuming that the bulk field satisfies the e.o.m. \eqref{eom}.

The correction to the first BOE coefficient, $c_\hD^{(1)}$, differs for different b.c.'s, and e.g. for DD it is given by
\begin{equation} \label{1st BOE coeff corr}
\begin{aligned}
c_\hD^{(1)} = \al \, \pi^2 \mathcal{A} + \frac{(N + 2)\la_*}{32\,\pi^2\th^2\ep} \left( m \left( 2 - \log\left(\frac{\pi}{\th}\right) - \sum_{k = 1}^{m - 1}\frac{1}{k} \right) - \frac{3}{2} \right) \ ,
\end{aligned}
\end{equation}
where we also see how this is related to the undetermined coefficient $\mathcal{A}$ from \eqref{Dyn fcn}. It is related in a similar way for the other b.c.'s.

\subsection{BOE coefficients}

Let us now study how we can find the other BOE coefficients that appear at $\mco(\ep)$ from \eqref{Dyn fcn}. As previously mentioned, by performing the summation in the differential operator in \eqref{Bulk-wedge corr} we find the conformal blocks, and in turn also the bootstrap equation \cite{Antunes:2021qpy}
\begin{equation} \label{2ptbeq}
\begin{aligned}
f(\vph) &= \sin^{\Dhhe_\al - \De_\ph}(\vph) \sum_{n} c_n \mathcal{G}_n \Big( \hat{\D}_n, \Dhh_\al, \tan\vph \Big) \\
&= \sin^{\Dhhe_\al - \De_\ph}(\th - \vph) \sum_{m} c'_{m} \mathcal{G}_m \Big( \hat{\D}_{m}', \Dhh_\al, \tan(\th - \vph) \Big) \ ,
\end{aligned}
\end{equation}
with the conformal blocks
\begin{equation}
\begin{aligned}
\mathcal{G}_n(\hat{\D}_n, \Dhh_\al, \e) &= \e^{\hat{\D}_n - \Dhhe_\al}\,{}_2F_1\left( \frac{\hD_n - \Dhh_\al}{2}, \frac{\hD_n - \Dhh_\al + 1}{2}, \Dh_n - \frac{d - 3}{2}, -\e^2 \right) \ .
\end{aligned}
\end{equation}
Note that the bootstrap equation \eqref{2ptbeq} is symmetric under $\vph \rightarrow \th - \vph$, which is the same as switching the two boundaries with each other: wall $\longleftrightarrow$ ramp. This explains why DN b.c.'s is the same as ND (Neumann on the wall and Dirichlet on the ramp).

In the $\ep$-expansion (\ref{eps exp}, \ref{eps exp 2}) we write
\begin{equation}
\begin{aligned}
\mathcal{G}_n ( \hat{\D}_n, \Dhh_\al, \e ) &= \mathcal{G}_n^{(0)} ( \hat{\D}_n^{(0)}, \Dhh_\al^{(0)}, \e ) + \ep\, \mathcal{G}_n^{(1)} ( \hat{\D}_n^{(0)}, \Dhh_\al^{(0)}, \hg_n, \hathat{\g}_\al, \e ) \ , \\
\mathcal{G}_n^{(0)} ( \hat{\D}_n^{(0)}, \Dhh_\al^{(0)}, \e ) &= \mathcal{G}_n ( \hat{\D}_n^{(0)}, \Dhh_\al^{(0)}, \e ) \ ,
\end{aligned}
\end{equation}
which gives us the wall block decomposition
\begin{equation} \label{wall block decomp free}
\begin{aligned}
f^{(0)}(\vph) &= \sin^{\al}\vph \sum_{n} c_{n}^{(0)}\, \mathcal{G}_n^{(0)} ( \hat{\D}_{n}^{(0)}, \Dhh_\al^{(0)}, \tan\vph ) \ , \\
\end{aligned}
\end{equation}
\begin{equation} \label{wall block decomp int}
\begin{aligned}
f^{(1)}(\vph) &= \sin^{\al}\vph\sum_{n} \bigg[ c_{n}^{(1)}\, \mathcal{G}_n^{(0)} ( \hat{\D}_n^{(0)}, \Dhh_\al^{(0)}, \tan\vph ) + \\
\eq + c_{n}^{(0)}\, \mathcal{G}_n^{(1)} ( \hat{\D}_{n}^{(0)}, \Dhh_\al^{(0)}, \hg_{0/1}, \hathat{\g}_\al, \tan\vph ) +  \\
\eq + (\hhg - \g_\ph)\,\log (\sin \varphi)\, c_{n}^{(0)}\, \mathcal{G}_n^{(0)} ( \hat{\D}_n^{(0)}, \Dhh_\al^{(0)}, \tan\vph ) \bigg] \ .
\end{aligned}
\end{equation}
The conformal block in the free theory is orthogonal\footnote{This orthogonality is found in the same way as in \cite{Bissi:2018mcq, Dey:2020jlc}.}
\begin{equation} \label{block orth}
\begin{aligned}
\oint_{|\e| = \tilde{\ep}} \frac{d\e}{2\,\pi\, i} \X(\hD_m^{(0)}, \Dhh_\al^{(0)}, \e)\mathcal{G}_n^{(0)}(\hD_n^{(0)}, \Dhh_\al^{(0)}, \e) = \d_{mn} \ .
\end{aligned}
\end{equation}
The integration is over a small circle with radius $0 < \tilde{\ep} \ll 1$, i.e. we consider the residue at $\e = 0$, and the orthogonality weight function is given by
\begin{equation*}
\begin{aligned}
\X(\hD_m^{(0)}, \Dhh_\al^{(0)}, \e) = \e^{\Dhhe_\al^{(0)} - \hD_m^{(0)} - 1}\,{}_2F_1 \left( \frac{\Dhh_\al^{(0)} - \hD_m^{(0)} - 1}{2}, \frac{\Dhh_\al^{(0)} - \hD_m^{(0)}}{2} - 1, \frac{d + 1}{2} - \hD_m^{(0)}, -\e^2 \right) \ .
\end{aligned}
\end{equation*}
If we apply this orthogonality relation to $f^{(0)}(\vph)$ in \eqref{wall block decomp free} we find
\begin{equation}
\begin{aligned}
c_n^{(0)} &= \oint_{|\vph| = \tilde{\ep}}\frac{d\vph}{2\,\pi\, i} \frac{\X(\hD_n^{(0)}, \Dhh_\a^{(0)}, \tan\vph)}{\cos^{2}\vph\,\sin^{\al}\vph} f^{(0)}(\vph) \ .
\end{aligned}
\end{equation}
For the ramp-channel BOE coefficients we exchange $\cos^{-2}\vph \rightarrow -\cos^{-2}(\th - \vph)$ and take the residue at $\vph = \th$. Using the result of $f^{(0)}$ from Sec. \ref{Sec: Free EOM} we find for the different b.c.'s
\begin{equation}
\begin{aligned}
&\text{DD:} \quad &c_n^{(0)} &= \frac{ \G({\Dhhe_\al^{(0)}}) }{ \pi^{\De_\ph^{(0)}} \th\, \G(\al) } \de_{n1} \ , \quad &c_n^{\prime(0)} &= \frac{ (-1)^{\al + 1}\G({\Dhhe_\al^{(0)}}) }{ \pi^{\De_\ph^{(0)}} \th\, \G(\al) } \de_{n1} \ , \\
&\text{NN:} \quad &c_n^{(0)} &= \frac{ \G({\Dhhe_\al^{(0)}}) }{ \pi^{\De_\ph^{(0)}} \th\, \al\, \G(\al) } \de_{n0} \ , \quad &c_n^{\prime(0)} &= \frac{ (-1)^{\al}\G({\Dhhe_\al^{(0)}}) }{ \pi^{\De_\ph^{(0)}} \th\,\al\, \G(\al) } \de_{n0} \ , \\
&\text{DN:} \quad &c_n^{(0)} &= \frac{ \G({\Dhhe_\al^{(0)}}) }{ \pi^{\De_\ph^{(0)}} \th\, \G(\al) } \de_{n1} \ , \quad &c_n^{\prime(0)} &= -\frac{ i^{2\al + 1}\G({\Dhhe_\al^{(0)}}) }{ \pi^{\De_\ph^{(0)}} \th\,\al\, \G(\al) } \de_{n0} \ ,
\end{aligned}
\end{equation}
where $m$ is the integer in $\al$. This means that at $\mco(\ep)$ we only need the $\ep$-expansion of the two conformal blocks with $n \in \{0, 1\}$, i.e. $\mathcal{G}_{0/1}^{(1)}$. In App. \ref{App: Block} we outline how this is done. Applying the orthogonality relation \eqref{block orth} to \eqref{wall block decomp int} gives us
\begin{equation}
\begin{aligned}
c_n^{(1)} &= \oint_{|\vph| = \tilde{\ep}}\frac{d\vph}{2\,\pi\, i} \frac{\X(\hD_n^{(0)}, \Dhh_\a^{(0)}, \tan\vph)}{\cos^{2}\vph\,\sin^{\al}\vph} \left[ f^{(1)}(\vph) - c_{0/1}^{(0)}\, \mathcal{G}^{(1)} ( \hat{\D}_{0/1}^{(0)}, \Dhh_\al^{(0)}, \hg_{0/1}, \hathat{\g}_\al, \tan\vph ) + \rig \\
\eq \lef - (\hhg - \g_\ph)\,\log (\sin \varphi)\, c_{0/1}^{(0)}\, \mathcal{G}^{(0)} ( \hat{\D}_{0/1}^{(0)}, \Dhh_\al^{(0)}, \tan\vph ) \right] \ .
\end{aligned}
\end{equation}
This integral is difficult to do in general, but it can be done for specific values of $n$. Only odd (Dirichlet) or even (Neumann) BOE coefficients (w.r.t. $n$) are non-zero, which agrees with the results in a BCFT \cite{Bissi:2018mcq}. Unfortunately we were not able to find the general form for the BOE coefficients, and we list the first few BOE coefficients for DD. The first one, $n = 1$, agrees with \eqref{1st BOE coeff corr}, and the next non-trivial one is given by
\begin{equation}
\begin{aligned}
c_5^{(1)} &= -\frac{(N + 2)m(\pi^4 - \th^4)\la_*}{9600\,\pi^2\th^6\ep} \ .
\end{aligned}
\end{equation}
Here $m$ is the same as in \eqref{edgescalingdim0}.


\section*{Acknowledgment}
AS is grateful for discussions with António Antunes at the early stages of this work. We thank Vladimír Procházka for illuminating discussions regarding conformal wedges and boundaries. We also thank António Antunes and Tobias Hansen for commenting an older version of the manuscript. This research received funding from the Knut and Alice Wallenberg Foundation grant KAW 2016.0129 and the VR grant 2018-04438.

\newpage

\appendix

\section{Computation of the propagator}
\label{app:prop}
In this appendix we give the details regarding the computation of the free bulk-bulk propagator reported in the main text (see also \cite{carslaw1910green,deutsch1979boundary} for previous discussions).
We need to compute
\begin{align*}
\left<\phi^i(\boldsymbol{x},\varphi,\rho)\phi^j(0,\varphi',\rho')\right> =& \frac{\de^{ij}}{2\theta}\sum_{m \in \Z} \int d^{d-1} k \, \frac{k_\rho}{2\, \omega (2\,\pi)^{d-3}} e^{-i\,\omega\, t+i\,  \boldsymbol{k} \cdot \boldsymbol{x}} J_{|\frac{\pi\, m}{\theta}|} \left(k_\rho\,\rho\right)J_{|\frac{\pi\, m}{\theta}|} \left(k_\rho\,\rho'\right) \times \\ 
&\times \left( e^{i \frac{\pi}{\theta}m (\varphi- \varphi')} \pm e^{i \frac{\pi}{\theta}m (\varphi+ \varphi')}  \right) \ ,
\end{align*}
where we promoted the fields to be invariant under $O(N)$. In the Euclidean formulation this may be rewritten as
\begin{equation}
\begin{split}
\left<\phi(\boldsymbol{x},\varphi,\rho)\phi(0,\varphi',\rho')\right> =   \frac{\pi\,\de^{ij}}{\theta} \sum_{m\in\Z} G_S^{(\frac{\pi\, m}{\theta})} (x_{||},\rho; x'_{||},\rho') \left( e^{i \frac{\pi}{\theta}m (\varphi- \varphi')} \pm e^{i \frac{\pi}{\theta}m (\varphi+ \varphi')}  \right) \ ,
\end{split}
\end{equation}
where we defined
\begin{equation}\label{eq:prop_mode_int}
\begin{split}
G_S^{(\nu)} (x_{||},\rho; x'_{||},\rho') \equiv   \int d^{d-3} {\boldsymbol{k}} \,dk_\rho\,d k_\tau \, \frac{k_\rho}{  (2\,\pi)^{d-1}} \frac{e^{-i\,k_\tau (\tau-\tau')+i\, {\boldsymbol{k} \cdot (\boldsymbol{x}- \boldsymbol{x}')}}}{k_\rho^2+\boldsymbol{k}^2 + k_\tau^2} J_{\nu} \left(k_\rho\rho\right)J_{\nu} \left(k_\rho\rho'\right) \ .
\end{split}
\end{equation}
By employing a Schwinger parametrisation
\begin{equation} \label{Sch param}
\frac{1}{A^n}= \int_0^\infty ds\frac{s^{n - 1}}{\G(n)}\, e^{-s\,A} \ , \quad A >0\ ,
\end{equation}
and performing the Gaussian integration over $\boldsymbol{k}$ and $k_\tau$, we obtain
\begin{equation}
\label{eq:int_step}
\begin{split}
G^{(\nu)}_S (x,x') &= \int_0^{+\infty} \,ds \int_0^{+\infty} d k_\rho \, \frac{k_\rho}{  2^{d-1} \pi^{d/2}} \frac{1}{s^{d/2-1}}e^{-\frac{\left(\boldsymbol{x}- \boldsymbol{x}' \right)^2+(\tau-\tau')^2}{4 \,s}} e^{-k_\rho^2 s} \\
\eq \times J_{\nu} \left(k_\rho\,\rho\right)J_{\nu} \left(k_\rho\,\rho'\right) \ .
\end{split}
\end{equation}
The integral over $k_\rho$ in \eqq{int_step} can be performed analytically (see for example \cite{book})
\begin{equation}
\label{eq:heat_kernel_prop_no_A}
\begin{split}
G^{(\nu)}_S (x,x') 
&= \frac{1}{2^{d}  \pi^{d/2}}  \int_0^\infty ds \, s^{\frac{d}{2}-2} e^{-s\frac{\rho^2+\rho'^2+\left(\boldsymbol{x}- \boldsymbol{x}' \right)^2}{4} }     I_{\nu} \left(\frac{s\,\rho\,\rho'}{2}\right) \ .
\end{split}
\end{equation}
At this point also the integral over $s$ in \eqq{heat_kernel_prop_no_A} can be done, and the result is expressed in equation \eqref{eq:G_S}.\footnote{We found it convenient to use the relation \begin{equation}
I_\beta(z)=e^{\mp i\beta \pi/2} J_\beta\left(z e^{\pm i \pi/2}\right) \ ,
\end{equation}  before computing the integral.}

A case of interest is when $\theta = \frac{\pi}{n}$. Indeed with this assumption, the sum  over $m$ simplifies and can be done. In this case the propagator reads
\begin{equation}
\label{eq:prop_int}
\begin{split}
\left<\phi(\boldsymbol{x},\varphi,\rho)\phi(0,\varphi',\rho')\right> =& \frac{n\,\de^{ij}}{2^{d-2}  \pi^{\frac{d}{2}}}  \int_0^\infty ds \, s^{\frac{d}{2}-2} e^{-s\frac{\rho^2+\rho'^2+\left(\boldsymbol{x}- \boldsymbol{x}' \right)^2}{4} } \times \\
&\times  \sum_{m\geq 0}{}^{\prime}  I_{m, n} \left(\frac{s\,\rho\,\rho'}{2}\right) \begin{cases}
\sin \left( m \, n\, \varphi \right) \sin \left( m \, n \, \varphi' \right)  \ ,&\quad \text{DD,} \\
\cos \left( m \, n \, \varphi \right) \cos \left( m \, n \, \varphi' \right) \ , &\quad \text{NN,}
\end{cases}
\end{split}
\end{equation}
where the sums can be expressed in terms of elementary functions as 
\begin{equation}
\begin{split}
&\sum_{m\geq 0} \sin ( n\, m\, \varphi) \sin ( n\, m\, \varphi') I_{m, n}(z) = \\
\eq \frac{1}{4\, n} \sum_{k = 0}^{n-1} \left\{ \exp\left[ z \cos \left( \frac{2\, k \,\pi + n(\varphi-\varphi')}{n} \right) \right] -  \exp\left[ z \cos \left( \frac{2\, k \,\pi + n(\varphi+\varphi')}{n} \right) \right] \right\} \  ,
\end{split}
\end{equation}
and
\begin{equation}
\begin{split}
&\sum_{m\geq 0}{}^{\prime} \cos ( n\, m\, \varphi) \cos ( n\, m\, \varphi') I_{m, n}(z) = \\
\eq \frac{1}{4 n} \sum_{k = 0}^{n-1} \left\{ \exp\left[ z \cos \left( \frac{2\, k\, \pi + n(\varphi-\varphi')}{n} \right) \right] +  \exp\left[ z \cos \left( \frac{2\, k\, \pi + n(\varphi+\varphi')}{n} \right) \right] \right\}  \ .
\end{split}
\end{equation}
By plugging those identities  back to \eqq{prop_int} and performing the integral over $s$ we obtain \eqq{phi-phi-images}.

\section{Feynman diagrams}
\label{app:feynman}

In this appendix we double check the results on the wedge anomalous dimensions by calculating Feynman diagrams. We do not find the full correlator in this approach, making the method in Sec. \ref{Sec: EOM} stronger. At the first order in perturbation theory, they can be extracted from the logarithmic contribution to the relevant Feynman diagram. In addition, since we are only interested in the leading order in $\epsilon$, we will just evaluate the relevant integrals at $d=4$ in Sec. \ref{sec:feynm_bulk} and at $d=3$ in Sec. \ref{app:wedge_anom_d3}. By conformal symmetry along the edge, the two-point function of defect primaries has the form
\begin{equation}
\label{eq:twoOO}
\langle \hathat O_{\alpha}^i (x_\para) \hathat O_{\beta}^j (y_\para)  \rangle = \de^{ij}\de_{\al\bet}\frac{C_{\hathatexp \Delta}}{|x_{||}|^{2\,{\hathatexp\Delta}}} \ ,
\end{equation}  
where $\hathat \Delta$ is the conformal dimension of the defect primary $\hathat O_\al$.  In a perturbative framework we may write
\begin{equation}
\label{eq:Delta_exp}
\hathat \Delta = \hathat \Delta^{(0)} + \ep\, \hathat \gamma + \mco(\ep^2) \ , \quad C_{\hathatexp \Delta} = C_{\hathatexp \Delta}^{(0)} + \ep \, C_{\hathatexp \Delta}^{(1)} + \mco(\ep^2) \ .
\end{equation}
By plugging this into \eqq{twoOO} and expanding in $\lambda$ we obtain
\begin{equation}
\langle \hathat O_{\alpha}^i (x_\para) \hathat O_{\beta}^j (y_\para)  \rangle =  \frac{\de^{ij}\de_{\al\bet}}{|s_{\para}|^{2\,{\hathatexp \Delta^{(0)}}}}  \left[ C_{\hathatexp \Delta}^{(0)} +  \left( C_{\hathatexp \Delta}^{(1)} - 2 C_{\hathatexp \Delta}^{(0)} \hathat \gamma^{(1)} \, \log |s_{||}| \right) \ep  \right] + \mathcal{O}(\ep^2) \ .
\end{equation}
This means that, at $\mco(\la)$, we can extract the anomalous dimension, $\hathat \gamma$, from the coefficient of the logarithmic divergence, $A_{\text{\tiny log}}$, as
\begin{equation} \label{Wedge anom dim formula}
\hathat \gamma = - \frac{A_{\text{\tiny log}}}{2 \,C_{\hathatexp \Delta}^{(0)} \ep}  \ .
\end{equation}

\subsection{Wedge anomalous dimension in $d = 4 - \ep$}
\label{sec:feynm_bulk}

The leading contribution to the anomalous dimension comes from the following integral
\begin{equation*} \hspace{-10px}
\begin{split}
\left< \hathat O^i_\a (x_{||})\hathat O^j_\al (y_\para)\right>_1 &= - \de^{ij}\frac{(N + 2)\la}{2}   \int_0^{+\infty} d\rho \, \rho \int_0^\theta d\varphi \int_{\R^{d - 2}} d^{d - 2} z_{||} \, \left< \hathat O^k_\a (x_{||}) \phi^k(z) \right>_0 \times \\ 
\eq\times \left< \ph^{k\, 2}(z) \right>_0 \left<  \ph^k(z) \hathat O^k_\al (y_\para)\right>_0  \ .
\end{split}
\end{equation*}
Here $k$ represents only one of the possible $N$ fields, and  in the integrand we have the correlators from the free theory reported in Sec. \ref{Sec:Free}. We are interested in the case when $\al > 0$. We find it convenient to rewrite the integral as
\begin{equation} \label{Feyn corr}
\begin{split}
\langle \hathat O^i_\a (x_{||})\hathat O^j_\al (y_\para)\rangle_1 &= - \de^{ij} \frac{(N + 2)\lambda}{64\, \pi^4 \, \theta^4}  \mathcal{I}(\th)\mathcal{J}(s_\para) \ , 
\end{split}
\end{equation}
where we defined
\begin{equation}
\begin{split}
\mathcal{I}(\th) &\equiv \int_0^\theta d\varphi \,  g_\al(\vph)h(\varphi) \ , \\
\mathcal{J}(s_{||}) &\equiv \int_0^{+\infty} d\rho  \int d^2 z_{||} \,\frac{1}{\rho^3} \left(\frac{\rho}{\rho ^2+\left(s_{||}+z_{||}\right)^2}\right)^{\a+1}  \left(\frac{\rho}{\rho ^2+z_{||}^2}\right)^{\a+1} \ .
\end{split}
\end{equation}
The angular integral can be done directly by introducing an angular cutoff $\vartheta$ s.t. $\vph \in (\vartheta, \th - \vartheta)$
\beq
\mathcal{I}(\th) =
\begin{cases}
\displaystyle{\frac{(\pi^2 -\theta^2 )  \theta}{2} -3\, \pi^2 m\, \theta}  \ ,  \quad & DD, \\
\displaystyle{\frac{2\, \theta^3}{3\,\pi\, \vartheta (\th - \vartheta)}  +\frac{(\pi^2 -\theta^2 )  \theta}{2} - 3 \,\pi^2 m \,\theta} \ ,  \quad & NN, \\
\displaystyle{\frac{3 \,\theta ^3}{\th - \vartheta} - \frac{\theta^3}{2} - \frac{\pi ^2 \theta  (12\,m + 7)}{4}} \ ,   \quad &  DN .
\end{cases}
\eeq
For the purpose of finding the wedge anomalous dimension we only need to care about the finite part of $\mathcal{I}(\th)$, and discard the non-universal power-law divergence. Note that the result corresponding to DD and NN are the same, while the DN case differs. The poles in $\vartheta$ or $\th - \vartheta$ corresponds to the $\hp$ exchange on its respective boundary for Neumann b.c.

We are left with analysing the function $\mathcal{J}(s_{||})$. After a Feynman parametrisation, and a shift $z_\para \rightarrow z_\para + u\,s_\para$
\begin{equation}
\begin{split}
\mathcal{J}(s_{||})  = & \frac{\Gamma({2(\al + 1)})}{\Gamma({\al + 1})^2} \int_0^{+\infty} d\rho  \int_0^{1} du  \int_{\R^{d - 2}} d^{d - 2} z_{||} \, \rho^{2\al -1} \frac{\left[u(1-u) \right]^{\al}}{\left[ \rho^2 + z^2_{||} + s_{||}^2 \, u (1-u)  \right]^{ 2(\al + 1) }} \ .
\end{split}
\end{equation}
The integral over $z_\para$ can now be done using a Schwinger parametrization \eqref{Sch param}
\begin{equation}
\begin{split}
\mathcal{J}(s_{||})  = \frac{\Gamma \left( 2\,\al + 1 \right)}{\Gamma({\al + 1})^2} \frac{\pi}{\left|s_{||}\right|^{2(2\,\al + 1)}}  \int_0^{+\infty} d\rho \int_0^{1} du  \, \rho^{2\,\al - 1} \frac{\left[u(1-u)\right]^{\al}}{\left(\frac{\rho^2}{ s_{||}^2} +  u (1-u)  \right)^{2\,\al + 1 }}  \ .
\end{split}
\end{equation}
At this point, if we perform the integral over $\rho$ first, we see that the result is finite without the introduction of any cutoff. However, the integral over $u$ turns out to be logarithmic divergent. To better expose this divergence we first rewrite the $\rho$-integral in terms of $\rho^2$ and then we introduce a cutoff $\varrho$ s.t. $\rho^2 \ge \varrho^2$. We get
\begin{equation*}
\begin{split}
\mathcal{J}(s_{||})  &= \frac{\Gamma \left( 2\,\al + 1 \right)}{\Gamma({\al + 1})^2} \frac{\pi}{2\left|s_{||}\right|^{2(2\,\al + 1)}}  \int_{\varrho^2}^{+\infty} d\rho^2 \int_0^{1} du  \, \left(\rho^2\right)^{\al-1} \frac{\left[u(1-u) \right]^{\al}}{\left(\frac{\rho^2}{ s_{||}^2} +  u (1-u)  \right)^{2\,\al + 1 }} \\
&= \frac{\Gamma \left( 2\,\al + 1 \right)}{\Gamma({\al + 1})^2} \frac{\pi}{2\left|s_{||}\right|^{2(2\,\al + 1)}}  \int_{0}^{+\infty} d\rho^2 \int_0^{1} du  \, \left(\rho^2\right)^{\al-1} \frac{\left[u(1-u) \right]^{\al}}{\left(\frac{\rho^2}{ s_{||}^2} + \frac{\varrho^2}{ s_{||}^2} +  u (1-u)  \right)^{2\,\al + 1 }} + ... \ .
\end{split}
\end{equation*}
Note that in the second step we did the shift $\rho^2 \rightarrow \rho^2 +\varrho^2$. In an expansion in the cutoff we can neglect the $\varrho$-contribution from the $\rho^2$-term of the numerator. Performing the integral over $\rho^2$ gives
\begin{equation}
\mathcal{J}(s_{||}) =\frac{\theta}{2\,m\left|s_{||}\right|^{2(\al + 1)}} \int_0^{1} du \frac{\left[u(1-u )\right]^{\al} }{\left(\frac{\varrho^2}{s_{||}^2}+ u ( 1 - u ) \right)^{\al + 1}} \ . 
\end{equation}
Thanks to the cutoff $\varrho$, now also the integral over $u$ is convergent, and it can be performed analytically. The result is a hypergeometric function which we do not report here. By expanding the result in the cutoff we find
\begin{equation}
\label{eq:I_exp}
\mathcal{J}(s_{||}) =  \frac{2\,\pi}{\a\left|x_{||}\right|^{2(\a+1)}}\left[  \log \left(\frac{\left|s_{||}\right|}{\varrho} \right) -\frac{1}{2\a} -\frac{\psi\left(\a \right)}{2}- \frac{\g_E}{2}   \right] + \mathcal{O}(\varrho) \  ,
\end{equation}
where $\ps(\al)$ is the digamma function, and $\g_E$ is the Euler-Mascheroni constant.
As discussed above, the correction to the conformal dimension can be found from the coefficient of the logarithmic term in the correlator \eqref{Feyn corr}. 

All and all, this gives
\begin{equation}
\begin{aligned}
C_{{\hathatexp \Delta}}^{(0)} &= \frac{1}{\pi \, \theta} \ , \quad A_{\text{\tiny log}} = \frac{(N + 2)\la}{96\,\pi^4\th^2}
\left\{ \begin{array}{l l}
\displaystyle{\frac{ \pi ^2 ( 6 \,m-1) + \theta ^2}{m}} \ , &\quad \text{DD, NN,} \\
\\
\displaystyle{\frac{\pi ^2   (12\,m + 7) + 2\,\theta^2}{2\,m + 1}} \ , &\quad \text{DN.}
\end{array} \right.
\end{aligned}
\end{equation}
The formula \eqref{Wedge anom dim formula} then reproduces the wedge anomalous dimensions in \eqref{Wedge anom dim}.

\subsection{Wedge anomalous dimension in $d = 3 - \ep$}
\label{app:wedge_anom_d3}

In this case we need to compute 
\begin{equation*} \hspace{-10px}
\begin{split}
\left< \hathat O^i_\a (x_{||})\hathat O^j_\al (y_\para)\right>_1  & = - \de^{ij}\frac{(N + 2)g_*}{2} \int_0^{+\infty} d\rho  \int_{\R^{d - 2}} d z_{||} \left< \hathat O^k_\a (x_{||}) \hat\phi^k(z) \right>_0  \left< \hat\phi^{2}(z)  \right>_0 \left<  \hat \phi^k(z) \hathat O^k_\a (y_\para)\right>_0 \ ,
\end{split}
\end{equation*}
%
where again $k$ represents only one of the $N$ possible fields.
Analogously to the computation done above, we need the boundary-wedge propagator and the one-point function of $\ph^2$ when $d=3$. Those correlators have been reported in (\ref{eq:bound_wedge}, \ref{eq:bdy_phi_sq}).

Thus, the integral to compute is 
\begin{equation*} 
\begin{aligned}
\begin{split}
\left< \hathat O^i_\a (x_{||})\hathat O^j_\al (y_\para)\right>_1   =  &- \de^{ij}\frac{(N + 2)g_* \k_n\Gamma\left({\alpha +\tfrac{1}{2}}\right)^2}{4\sqrt{2}\,\pi^4 \theta^2 \Gamma({\alpha +1})^2} \times \\
\eq \times \int_0^{+\infty} d\rho  \int d z_{||} \, \rho^{2\a -1}  \left( \frac{1}{\rho^2 + (s_{||}+z_{||})^2} \right)^{\a + \frac{1}{2}} \left( \frac{1}{\rho^2 + z^2_{||}} \right)^{\a + \frac{1}{2}}  \ .
\end{split}
\end{aligned}
\end{equation*}
By doing the same steps as in the case of Sec. \ref{sec:feynm_bulk} we obtain
\beq
\left< \hathat O^i_\a (x_{||})\hathat O^j_\al (y_\para)\right>_1  = \de^{ij}  \frac{ (N + 2)g_* \kappa_n \Gamma({2\, \alpha}) 
}{2^{2\, \alpha +\frac{3}{2}} \pi\, \theta^2 \Gamma({\alpha +1})^2s_{||}^{2\, \alpha + 1 }} \left[H_{\alpha -\frac{1}{2}}-2 \log \left(\frac{|s_{||}|}{\varrho} \right)\right]   + \mathcal{O}(\varrho) \ ,
\eeq
where $H_x$ is the analytic continuation of the harmonic number to real values. 
From the above equation we can extract the anomalous dimension, which gives us \eqref{3d wedge anom dim}.\footnote{Had we added the same interaction on both of the boundaries, there is an extra factor of $2$ in the wedge anomalous dimension.}

\section{Expansion of the conformal blocks} \label{App: Block}

In this appendix we explain how the conformal blocks from the free theory can be expanded in $\ep$. For Dirichlet b.c. we wish to expand
\begin{equation}
\begin{aligned}
\mathcal{G}_0(\hD_0, \Dhh_\al, z) &\propto {}_2F_1\left( -\frac{\al}{2} + a\,\ep, \frac{1 - \al}{2} + a\,\ep, \frac{1}{2} + b\,\ep, z \right) \ ,
\end{aligned}
\end{equation}
and for Neumann b.c.
\begin{equation}
\begin{aligned}
\mathcal{G}_1(\hD_0, \Dhh_\al, z) &\propto {}_2F_1\left( \frac{1 - \al}{2} + a\,\ep, 1 - \frac{\al}{2} + a\,\ep, \frac{3}{2} + b\,\ep, z \right) \ .
\end{aligned}
\end{equation}
The coefficients $a$ and $b$ are given by
\begin{equation}
\begin{aligned}
a &= \frac{\hg - \hhg}{2} \ , \quad b = \hg \ .
\end{aligned}
\end{equation}
These two blocks satisfy the relation
\begin{equation}
\begin{aligned}
{}_2F_1(a, b, c, z) &= (\sqrt{z} + 1)^{-2\,a} {}_2F_1 \left( 2\,a, c - \frac{1}{2}, 2\,c - 1, \frac{2\sqrt{z}}{\sqrt{z} + 1} \right) \ ,
\end{aligned}
\end{equation}
which holds if $b = a + \frac{1}{2}$. Applying this relation, and for the Dirichlet block also the following one of Kummer's relations
\begin{equation}
\begin{aligned}
{}_2F_1(a, b, c, z) &= \frac{ \G({1 - a})\G({1 - b})\G(c) }{ \G({2 - c})\G({c - a})\G({c - b}) } \frac{{}_2F_1(a - c + 1, b - c + 1, 2 - c, z)}{z^{c - 1}} + \\
\eq + \frac{ \G({1 - a})\G({1 - b}) }{ \G({1 - c})\G({c - a - b + 1}) } \frac{ {}_2F_1(c - a, c - b, c - a - b + 1, z) }{ (1 - z)^{a + b - c} } \ ,
\end{aligned}
\end{equation}
we find ${}_2F_1(a, b, c, z)$'s in the conformal blocks that all satisfy $c > b > \mco(\ep)$. This means that we can use the following real-line integral representation (on all of the ${}_2F_1$'s)\footnote{This integral is convergent for $b > a > 0$.}
\begin{equation}
\begin{aligned}
{}_2F_1(a, b, c, z) &= \frac{ \G(c) }{ \G(b)\G({c - b}) } \int_0^1dt \frac{ t^{b - 1} (1 - t)^{c - b - 1} }{ (1 - t\,z)^a }  \ .
\end{aligned}
\end{equation}
The integrand can then be expanded in $\ep$. This gives us integrals over different kinds of logarithms at $\mco(\ep)$, which we performed using the 'Rubi' package for Mathematica \cite{Rich2018}. We have then found the $\ep$-expansions of the ${}_2F_1$ in the blocks\footnote{These can also be found in the attached Mathematica file.}
\begin{equation}
\begin{aligned}
\mathcal{G}_0 &\propto \frac{\left(1-\sqrt{z}\right)^k+\left(1 + \sqrt{z}\right)^k}{2} + \epsilon  \left\{ (a-c)\frac{ \left(1 + \sqrt{z}\right)^k-\left(1-\sqrt{z}\right)^k}{k} + \rig \\
\eq + (c-a) \left(1 + \sqrt{z}\right)^k \log \left(1 + \sqrt{z}\right)-a \left(1-\sqrt{z}\right)^k \log \left(1-\sqrt{z}\right) + \\
\eq -c
\left(\sqrt{z}+1\right)^k \log \left(2 \sqrt{z}\right) + (c-a) H_{k-1} \left[\left(1-\sqrt{z}\right)^k-\left(1 + \sqrt{z}\right)^k\right] + \\
\eq + H_k \left[(a-c) \left(1-\sqrt{z}\right)^k-a \left(\sqrt{z}+1\right)^k\right] + \\
\eq -\frac{c \left(1-\sqrt{z}\right)^{k+1} }{(k + 1)(1 + \sqrt{z})}\, _2F_1\left(1,k+1;k+2;\frac{1 - \sqrt{z}}{1 + \sqrt{z}}\right) + \\
\eq + \frac{c k \sqrt{z}}{1 + \sqrt{z}}\left[\left(1 + \sqrt{z}\right)^k \, _3F_2\left(1,1,1-k;2,2;\frac{2\sqrt{z}}{\sqrt{z} + 1}\right) + \rig \\
\eq \lef\lef -\left(1-\sqrt{z}\right)^k \,
_3F_2\left(1,1,1-k;2,2;\frac{2\sqrt{z}}{\sqrt{z}-1}\right)\right] \right\} \ ,
\end{aligned}
\end{equation}
\vspace{1px}
\begin{equation}
\begin{aligned}
\mathcal{G}_1 &\propto \frac{\left(1 + \sqrt{z}\right)^k-\left(1-\sqrt{z}\right)^k}{2 k \sqrt{z}} + \ep \left\{ (a+c k)\frac{\left(1 + \sqrt{z}\right)^k - \left(1-\sqrt{z}\right)^k}{k^2 \sqrt{z}} + \rig \\
\eq + \frac{a \left(1-\sqrt{z}\right)^k \log \left(1-\sqrt{z}\right)}{k \sqrt{z}}-\frac{a \left(1 + \sqrt{z}\right)^k \log
	\left(1 + \sqrt{z}\right)}{k \sqrt{z}} + \\
\eq -c \left[ \left(1-\sqrt{z}\right)^{k-1} \, _3F_2\left(1,1,1-k;2,2;\frac{2 \sqrt{z}}{\sqrt{z}-1}\right) + \rig \\
\eq \lef\lef + \left(1 + \sqrt{z}\right)^{k-1} \,
_3F_2\left(1,1,1-k;2,2;\frac{2 \sqrt{z}}{1 + \sqrt{z}}\right) \right] \right\} \ .
\end{aligned}
\end{equation}

\bibliographystyle{JHEP}
\bibliography{paper}

\providecommand{\href}[2]{#2}\begingroup\raggedright\begin{thebibliography}{10}

\bibitem{deutsch1979boundary}
D.~Deutsch and P.~Candelas, \emph{Boundary effects in quantum field theory}, {\emph{Physical Review D} {\bfseries 20} (1979) 3063}.

\bibitem{C1983}
J.~L. Cardy, \emph{Critical behaviour at an edge}, \href{http://dx.doi.org/10.1088/0305-4470/16/15/026}{\emph{Journal of Physics A: Mathematical and General} {\bfseries 16} (oct, 1983) 3617--3628}.

\bibitem{GT1984}
A.~J. Guttmann and G.~M. Torrie, \emph{Critical behaviour at an edge for the {SAW} and ising model}, \href{http://dx.doi.org/10.1088/0305-4470/17/18/023}{\emph{Journal of Physics A: Mathematical and General} {\bfseries 17} (dec, 1984) 3539--3552}.

\bibitem{barber1984magnetization}
M.~N. Barber, I.~Peschel and P.~A. Pearce, \emph{Magnetization at corners in two-dimensional ising models}, {\emph{Journal of statistical physics} {\bfseries 37} (1984) 497--527}.

\bibitem{CR1984}
J.~L. Cardy and S.~Redner, \emph{Conformal invariance and self-avoiding walks in restricted geometries}, \href{http://dx.doi.org/10.1088/0305-4470/17/17/005}{\emph{Journal of Physics A: Mathematical and General} {\bfseries 17} (dec, 1984) L933--L938}.

\bibitem{kaiser1989surface}
C.~Kaiser and I.~Peschel, \emph{Surface and corner magnetizations in the two-dimensional ising model}, {\emph{Journal of statistical physics} {\bfseries 54} (1989) 567--579}.

\bibitem{pleimling1998critical}
M.~Pleimling and W.~Selke, \emph{Critical phenomena at edges and corners}, {\emph{The European Physical Journal B-Condensed Matter and Complex Systems} {\bfseries 5} (1998) 805--810}.

\bibitem{dowker1987vacuum}
J.~S. Dowker, \emph{Vacuum averages for arbitrary spin around a cosmic string}, {\emph{Physical Review D} {\bfseries 36} (1987) 3742}.

\bibitem{de1989classical}
P.~de~Sousa~Gerbert and R.~Jackiw, \emph{Classical and quantum scattering on a spinning cone}, {\emph{Communications in Mathematical Physics} {\bfseries 124} (1989) 229--260}.

\bibitem{alford1989aharonov}
M.~G. Alford and F.~Wilczek, \emph{Aharonov-bohm interaction of cosmic strings with matter}, {\emph{Physical Review Letters} {\bfseries 62} (1989) 1071}.

\bibitem{alford1989enhanced}
M.~G. Alford, J.~March-Russell and F.~Wilczek, \emph{Enhanced baryon number violation due to cosmic strings}, {\emph{Nuclear Physics B} {\bfseries 328} (1989) 140--158}.

\bibitem{Kibble_1976}
T.~W.~B. Kibble, \emph{Topology of cosmic domains and strings}, \href{http://dx.doi.org/10.1088/0305-4470/9/8/029}{\emph{Journal of Physics A: Mathematical and General} {\bfseries 9} (aug, 1976) 1387--1398}.

\bibitem{Copeland:2003bj}
E.~J. Copeland, R.~C. Myers and J.~Polchinski, \emph{{Cosmic F and D strings}}, \href{http://dx.doi.org/10.1088/1126-6708/2004/06/013}{\emph{JHEP} {\bfseries 06} (2004) 013}, [\href{https://arxiv.org/abs/hep-th/0312067}{{\ttfamily hep-th/0312067}}].

\bibitem{Callan:1994py}
C.~G. Callan, Jr. and F.~Wilczek, \emph{{On geometric entropy}}, \href{http://dx.doi.org/10.1016/0370-2693(94)91007-3}{\emph{Phys. Lett. B} {\bfseries 333} (1994) 55--61}, [\href{https://arxiv.org/abs/hep-th/9401072}{{\ttfamily hep-th/9401072}}].

\bibitem{Solodukhin:1994yz}
S.~N. Solodukhin, \emph{{The Conical singularity and quantum corrections to entropy of black hole}}, \href{http://dx.doi.org/10.1103/PhysRevD.51.609}{\emph{Phys. Rev. D} {\bfseries 51} (1995) 609--617}, [\href{https://arxiv.org/abs/hep-th/9407001}{{\ttfamily hep-th/9407001}}].

\bibitem{Rattazzi:2008pe}
R.~Rattazzi, V.~S. Rychkov, E.~Tonni and A.~Vichi, \emph{{Bounding scalar operator dimensions in 4D CFT}}, \href{http://dx.doi.org/10.1088/1126-6708/2008/12/031}{\emph{JHEP} {\bfseries 12} (2008) 031}, [\href{https://arxiv.org/abs/0807.0004}{{\ttfamily 0807.0004}}].

\bibitem{Poland:2018epd}
D.~Poland, S.~Rychkov and A.~Vichi, \emph{{The Conformal Bootstrap: Theory, Numerical Techniques, and Applications}}, \href{http://dx.doi.org/10.1103/RevModPhys.91.015002}{\emph{Rev. Mod. Phys.} {\bfseries 91} (2019) 015002}, [\href{https://arxiv.org/abs/1805.04405}{{\ttfamily 1805.04405}}].

\bibitem{Bissi:2022mrs}
A.~Bissi, A.~Sinha and X.~Zhou, \emph{{Selected Topics in Analytic Conformal Bootstrap: A Guided Journey}},  \href{https://arxiv.org/abs/2202.08475}{{\ttfamily 2202.08475}}.

\bibitem{Antunes:2021qpy}
A.~Antunes, \emph{{Conformal bootstrap near the edge}}, \href{http://dx.doi.org/10.1007/JHEP10(2021)057}{\emph{JHEP} {\bfseries 10} (2021) 057}, [\href{https://arxiv.org/abs/2103.03132}{{\ttfamily 2103.03132}}].

\bibitem{Liendo:2012hy}
P.~Liendo, L.~Rastelli and B.~C. van Rees, \emph{{The Bootstrap Program for Boundary CFT$_d$}}, \href{http://dx.doi.org/10.1007/JHEP07(2013)113}{\emph{JHEP} {\bfseries 07} (2013) 113}, [\href{https://arxiv.org/abs/1210.4258}{{\ttfamily 1210.4258}}].

\bibitem{Bissi:2019kkx}
A.~Bissi, P.~Dey and T.~Hansen, \emph{{Dispersion Relation for CFT Four-Point Functions}}, \href{http://dx.doi.org/10.1007/JHEP04(2020)092}{\emph{JHEP} {\bfseries 04} (2020) 092}, [\href{https://arxiv.org/abs/1910.04661}{{\ttfamily 1910.04661}}].

\bibitem{Bissi:2021spj}
A.~Bissi, P.~Dey and G.~Fardelli, \emph{{Two Applications of the Analytic Conformal Bootstrap: A Quick Tour Guide}}, \href{http://dx.doi.org/10.3390/universe7070247}{\emph{Universe} {\bfseries 7} (2021) 247}, [\href{https://arxiv.org/abs/2107.10097}{{\ttfamily 2107.10097}}].

\bibitem{Bissi:2018mcq}
A.~Bissi, T.~Hansen and A.~S\"oderberg, \emph{{Analytic Bootstrap for Boundary CFT}}, \href{http://dx.doi.org/10.1007/JHEP01(2019)010}{\emph{JHEP} {\bfseries 01} (2019) 010}, [\href{https://arxiv.org/abs/1808.08155}{{\ttfamily 1808.08155}}].

\bibitem{Dey:2020jlc}
P.~Dey and A.~S\"oderberg, \emph{{On analytic bootstrap for interface and boundary CFT}}, \href{http://dx.doi.org/10.1007/JHEP07(2021)013}{\emph{JHEP} {\bfseries 07} (2021) 013}, [\href{https://arxiv.org/abs/2012.11344}{{\ttfamily 2012.11344}}].

\bibitem{Bianchi:2022ppi}
L.~Bianchi and D.~Bonomi, \emph{{Conformal dispersion relations for defects and boundaries}},  \href{https://arxiv.org/abs/2205.09775}{{\ttfamily 2205.09775}}.

\bibitem{Barrat:2022psm}
J.~Barrat, A.~Gimenez-Grau and P.~Liendo, \emph{{A dispersion relation for defect CFT}},  \href{https://arxiv.org/abs/2205.09765}{{\ttfamily 2205.09765}}.

\bibitem{Sinha:2022sdo}
A.~Sinha, \emph{{Dispersion relations and knot theory}},  \href{https://arxiv.org/abs/2204.13986}{{\ttfamily 2204.13986}}.

\bibitem{Maldacena:1997re}
J.~M. Maldacena, \emph{{The Large N limit of superconformal field theories and supergravity}}, \href{http://dx.doi.org/10.1023/A:1026654312961}{\emph{Adv. Theor. Math. Phys.} {\bfseries 2} (1998) 231--252}, [\href{https://arxiv.org/abs/hep-th/9711200}{{\ttfamily hep-th/9711200}}].

\bibitem{Witten:1998qj}
E.~Witten, \emph{{Anti-de Sitter space and holography}}, \href{http://dx.doi.org/10.4310/ATMP.1998.v2.n2.a2}{\emph{Adv. Theor. Math. Phys.} {\bfseries 2} (1998) 253--291}, [\href{https://arxiv.org/abs/hep-th/9802150}{{\ttfamily hep-th/9802150}}].

\bibitem{Gubser:1998bc}
S.~S. Gubser, I.~R. Klebanov and A.~M. Polyakov, \emph{{Gauge theory correlators from noncritical string theory}}, \href{http://dx.doi.org/10.1016/S0370-2693(98)00377-3}{\emph{Phys. Lett. B} {\bfseries 428} (1998) 105--114}, [\href{https://arxiv.org/abs/hep-th/9802109}{{\ttfamily hep-th/9802109}}].

\bibitem{Takayanagi:2011zk}
T.~Takayanagi, \emph{{Holographic Dual of BCFT}}, \href{http://dx.doi.org/10.1103/PhysRevLett.107.101602}{\emph{Phys. Rev. Lett.} {\bfseries 107} (2011) 101602}, [\href{https://arxiv.org/abs/1105.5165}{{\ttfamily 1105.5165}}].

\bibitem{Fujita:2011fp}
M.~Fujita, T.~Takayanagi and E.~Tonni, \emph{{Aspects of AdS/BCFT}}, \href{http://dx.doi.org/10.1007/JHEP11(2011)043}{\emph{JHEP} {\bfseries 11} (2011) 043}, [\href{https://arxiv.org/abs/1108.5152}{{\ttfamily 1108.5152}}].

\bibitem{Geng:2021iyq}
H.~Geng, S.~L\"ust, R.~K. Mishra and D.~Wakeham, \emph{{Holographic BCFTs and Communicating Black Holes}}, \href{http://dx.doi.org/10.1007/JHEP08(2021)003}{\emph{jhep} {\bfseries 08} (2021) 003}, [\href{https://arxiv.org/abs/2104.07039}{{\ttfamily 2104.07039}}].

\bibitem{Ryu:2006bv}
S.~Ryu and T.~Takayanagi, \emph{{Holographic derivation of entanglement entropy from AdS/CFT}}, \href{http://dx.doi.org/10.1103/PhysRevLett.96.181602}{\emph{Phys. Rev. Lett.} {\bfseries 96} (2006) 181602}, [\href{https://arxiv.org/abs/hep-th/0603001}{{\ttfamily hep-th/0603001}}].

\bibitem{Ryu:2006ef}
S.~Ryu and T.~Takayanagi, \emph{{Aspects of Holographic Entanglement Entropy}}, \href{http://dx.doi.org/10.1088/1126-6708/2006/08/045}{\emph{JHEP} {\bfseries 08} (2006) 045}, [\href{https://arxiv.org/abs/hep-th/0605073}{{\ttfamily hep-th/0605073}}].

\bibitem{FarajiAstaneh:2017hqv}
A.~Faraji~Astaneh, C.~Berthiere, D.~Fursaev and S.~N. Solodukhin, \emph{{Holographic calculation of entanglement entropy in the presence of boundaries}}, \href{http://dx.doi.org/10.1103/PhysRevD.95.106013}{\emph{Phys. Rev. D} {\bfseries 95} (2017) 106013}, [\href{https://arxiv.org/abs/1703.04186}{{\ttfamily 1703.04186}}].

\bibitem{Chu:2017aab}
C.-S. Chu, R.-X. Miao and W.-Z. Guo, \emph{{On New Proposal for Holographic BCFT}}, \href{http://dx.doi.org/10.1007/JHEP04(2017)089}{\emph{JHEP} {\bfseries 04} (2017) 089}, [\href{https://arxiv.org/abs/1701.07202}{{\ttfamily 1701.07202}}].

\bibitem{Seminara:2017hhh}
D.~Seminara, J.~Sisti and E.~Tonni, \emph{{Corner contributions to holographic entanglement entropy in AdS$_{4}$/BCFT$_{3}$}}, \href{http://dx.doi.org/10.1007/JHEP11(2017)076}{\emph{JHEP} {\bfseries 11} (2017) 076}, [\href{https://arxiv.org/abs/1708.05080}{{\ttfamily 1708.05080}}].

\bibitem{Seminara:2018pmr}
D.~Seminara, J.~Sisti and E.~Tonni, \emph{{Holographic entanglement entropy in AdS$_{4}$/BCFT$_{3}$ and the Willmore functional}}, \href{http://dx.doi.org/10.1007/JHEP08(2018)164}{\emph{JHEP} {\bfseries 08} (2018) 164}, [\href{https://arxiv.org/abs/1805.11551}{{\ttfamily 1805.11551}}].

\bibitem{Hertzberg:2010uv}
M.~P. Hertzberg and F.~Wilczek, \emph{{Some Calculable Contributions to Entanglement Entropy}}, \href{http://dx.doi.org/10.1103/PhysRevLett.106.050404}{\emph{Phys. Rev. Lett.} {\bfseries 106} (2011) 050404}, [\href{https://arxiv.org/abs/1007.0993}{{\ttfamily 1007.0993}}].

\bibitem{Berthiere:2018ouo}
C.~Berthiere, \emph{{Boundary-corner entanglement for free bosons}}, \href{http://dx.doi.org/10.1103/PhysRevB.99.165113}{\emph{Phys. Rev. B} {\bfseries 99} (2019) 165113}, [\href{https://arxiv.org/abs/1811.12875}{{\ttfamily 1811.12875}}].

\bibitem{Nozaki:2012qd}
M.~Nozaki, T.~Takayanagi and T.~Ugajin, \emph{{Central Charges for BCFTs and Holography}}, \href{http://dx.doi.org/10.1007/JHEP06(2012)066}{\emph{JHEP} {\bfseries 06} (2012) 066}, [\href{https://arxiv.org/abs/1205.1573}{{\ttfamily 1205.1573}}].

\bibitem{Solodukhin:2015eca}
S.~N. Solodukhin, \emph{{Boundary terms of conformal anomaly}}, \href{http://dx.doi.org/10.1016/j.physletb.2015.11.036}{\emph{Phys. Lett. B} {\bfseries 752} (2016) 131--134}, [\href{https://arxiv.org/abs/1510.04566}{{\ttfamily 1510.04566}}].

\bibitem{Fursaev:2015wpa}
D.~Fursaev, \emph{{Conformal anomalies of CFT\textquoteright{}s with boundaries}}, \href{http://dx.doi.org/10.1007/JHEP12(2015)112}{\emph{JHEP} {\bfseries 12} (2015) 112}, [\href{https://arxiv.org/abs/1510.01427}{{\ttfamily 1510.01427}}].

\bibitem{FarajiAstaneh:2021foi}
A.~Faraji~Astaneh and S.~N. Solodukhin, \emph{{Boundary conformal invariants and the conformal anomaly in five dimensions}}, \href{http://dx.doi.org/10.1016/j.physletb.2021.136282}{\emph{Phys. Lett. B} {\bfseries 816} (2021) 136282}, [\href{https://arxiv.org/abs/2102.07661}{{\ttfamily 2102.07661}}].

\bibitem{Chalabi:2021jud}
A.~Chalabi, C.~P. Herzog, A.~O'Bannon, B.~Robinson and J.~Sisti, \emph{{Weyl anomalies of four dimensional conformal boundaries and defects}}, \href{http://dx.doi.org/10.1007/JHEP02(2022)166}{\emph{JHEP} {\bfseries 02} (2022) 166}, [\href{https://arxiv.org/abs/2111.14713}{{\ttfamily 2111.14713}}].

\bibitem{Friedan:2003yc}
D.~Friedan and A.~Konechny, \emph{{On the boundary entropy of one-dimensional quantum systems at low temperature}}, \href{http://dx.doi.org/10.1103/PhysRevLett.93.030402}{\emph{Phys. Rev. Lett.} {\bfseries 93} (2004) 030402}, [\href{https://arxiv.org/abs/hep-th/0312197}{{\ttfamily hep-th/0312197}}].

\bibitem{Jensen:2015swa}
K.~Jensen and A.~O'Bannon, \emph{{Constraint on Defect and Boundary Renormalization Group Flows}}, \href{http://dx.doi.org/10.1103/PhysRevLett.116.091601}{\emph{Phys. Rev. Lett.} {\bfseries 116} (2016) 091601}, [\href{https://arxiv.org/abs/1509.02160}{{\ttfamily 1509.02160}}].

\bibitem{Casini:2018nym}
H.~Casini, I.~Salazar~Landea and G.~Torroba, \emph{{Irreversibility in quantum field theories with boundaries}}, \href{http://dx.doi.org/10.1007/JHEP04(2019)166}{\emph{JHEP} {\bfseries 04} (2019) 166}, [\href{https://arxiv.org/abs/1812.08183}{{\ttfamily 1812.08183}}].

\bibitem{Cuomo:2021rkm}
G.~Cuomo, Z.~Komargodski and A.~Raviv-Moshe, \emph{{Renormalization Group Flows on Line Defects}}, \href{http://dx.doi.org/10.1103/PhysRevLett.128.021603}{\emph{Phys. Rev. Lett.} {\bfseries 128} (2022) 021603}, [\href{https://arxiv.org/abs/2108.01117}{{\ttfamily 2108.01117}}].

\bibitem{Wang:2021mdq}
Y.~Wang, \emph{{Defect a-theorem and a-maximization}}, \href{http://dx.doi.org/10.1007/JHEP02(2022)061}{\emph{JHEP} {\bfseries 02} (2022) 061}, [\href{https://arxiv.org/abs/2101.12648}{{\ttfamily 2101.12648}}].

\bibitem{Kobayashi:2018lil}
N.~Kobayashi, T.~Nishioka, Y.~Sato and K.~Watanabe, \emph{{Towards a $C$-theorem in defect CFT}}, \href{http://dx.doi.org/10.1007/JHEP01(2019)039}{\emph{JHEP} {\bfseries 01} (2019) 039}, [\href{https://arxiv.org/abs/1810.06995}{{\ttfamily 1810.06995}}].

\bibitem{Huang:2016rol}
K.-W. Huang, \emph{{Boundary Anomalies and Correlation Functions}}, \href{http://dx.doi.org/10.1007/JHEP08(2016)013}{\emph{JHEP} {\bfseries 08} (2016) 013}, [\href{https://arxiv.org/abs/1604.02138}{{\ttfamily 1604.02138}}].

\bibitem{Herzog:2017xha}
C.~P. Herzog and K.-W. Huang, \emph{{Boundary Conformal Field Theory and a Boundary Central Charge}}, \href{http://dx.doi.org/10.1007/JHEP10(2017)189}{\emph{JHEP} {\bfseries 10} (2017) 189}, [\href{https://arxiv.org/abs/1707.06224}{{\ttfamily 1707.06224}}].

\bibitem{Herzog:2017kkj}
C.~Herzog, K.-W. Huang and K.~Jensen, \emph{{Displacement Operators and Constraints on Boundary Central Charges}}, \href{http://dx.doi.org/10.1103/PhysRevLett.120.021601}{\emph{Phys. Rev. Lett.} {\bfseries 120} (2018) 021601}, [\href{https://arxiv.org/abs/1709.07431}{{\ttfamily 1709.07431}}].

\bibitem{Miao:2017aba}
R.-X. Miao and C.-S. Chu, \emph{{Universality for Shape Dependence of Casimir Effects from Weyl Anomaly}}, \href{http://dx.doi.org/10.1007/JHEP03(2018)046}{\emph{JHEP} {\bfseries 03} (2018) 046}, [\href{https://arxiv.org/abs/1706.09652}{{\ttfamily 1706.09652}}].

\bibitem{Jensen:2018rxu}
K.~Jensen, A.~O'Bannon, B.~Robinson and R.~Rodgers, \emph{{From the Weyl Anomaly to Entropy of Two-Dimensional Boundaries and Defects}}, \href{http://dx.doi.org/10.1103/PhysRevLett.122.241602}{\emph{Phys. Rev. Lett.} {\bfseries 122} (2019) 241602}, [\href{https://arxiv.org/abs/1812.08745}{{\ttfamily 1812.08745}}].

\bibitem{Prochazka:2018bpb}
V.~Prochazka, \emph{{The Conformal Anomaly in bCFT from Momentum Space Perspective}}, \href{http://dx.doi.org/10.1007/JHEP10(2018)170}{\emph{JHEP} {\bfseries 10} (2018) 170}, [\href{https://arxiv.org/abs/1804.01974}{{\ttfamily 1804.01974}}].

\bibitem{Lauria:2020emq}
E.~Lauria, P.~Liendo, B.~C. Van~Rees and X.~Zhao, \emph{{Line and surface defects for the free scalar field}}, \href{http://dx.doi.org/10.1007/JHEP01(2021)060}{\emph{JHEP} {\bfseries 01} (2021) 060}, [\href{https://arxiv.org/abs/2005.02413}{{\ttfamily 2005.02413}}].

\bibitem{Gaiotto:2013nva}
D.~Gaiotto, D.~Mazac and M.~F. Paulos, \emph{{Bootstrapping the 3d Ising twist defect}}, \href{http://dx.doi.org/10.1007/JHEP03(2014)100}{\emph{JHEP} {\bfseries 03} (2014) 100}, [\href{https://arxiv.org/abs/1310.5078}{{\ttfamily 1310.5078}}].

\bibitem{Bianchi:2021snj}
L.~Bianchi, A.~Chalabi, V.~Proch\'azka, B.~Robinson and J.~Sisti, \emph{{Monodromy defects in free field theories}}, \href{http://dx.doi.org/10.1007/JHEP08(2021)013}{\emph{JHEP} {\bfseries 08} (2021) 013}, [\href{https://arxiv.org/abs/2104.01220}{{\ttfamily 2104.01220}}].

\bibitem{sommerfeld1897vieldeutige}
A.~Sommerfeld, \emph{{\"U}ber vieldeutige potenziale in raum}, {\emph{Proc. London Math. Soc.} {\bfseries 28} (1897) 395}.

\bibitem{diehl1986field}
H.~Diehl, \emph{Field-theoretic approach to critical behaviour at surfaces}.
\newblock Academic Press, 1986.

\bibitem{Dowker:1977zj}
J.~S. Dowker, \emph{{Quantum Field Theory on a Cone}}, \href{http://dx.doi.org/10.1088/0305-4470/10/1/023}{\emph{J. Phys. A} {\bfseries 10} (1977) 115--124}.

\bibitem{Bianchi:2015liz}
L.~Bianchi, M.~Meineri, R.~C. Myers and M.~Smolkin, \emph{{R\'enyi entropy and conformal defects}}, \href{http://dx.doi.org/10.1007/JHEP07(2016)076}{\emph{JHEP} {\bfseries 07} (2016) 076}, [\href{https://arxiv.org/abs/1511.06713}{{\ttfamily 1511.06713}}].

\bibitem{McAvity:1995zd}
D.~M. McAvity and H.~Osborn, \emph{{Conformal field theories near a boundary in general dimensions}}, \href{http://dx.doi.org/10.1016/0550-3213(95)00476-9}{\emph{Nucl. Phys. B} {\bfseries 455} (1995) 522--576}, [\href{https://arxiv.org/abs/cond-mat/9505127}{{\ttfamily cond-mat/9505127}}].

\bibitem{Dowker_1978}
J.~S. Dowker and G.~Kennedy, \emph{Finite temperature and boundary effects in static space-times}, \href{http://dx.doi.org/10.1088/0305-4470/11/5/020}{\emph{Journal of Physics A: Mathematical and General} {\bfseries 11} (may, 1978) 895--920}.

\bibitem{Dowker:2004mq}
J.~S. Dowker, \emph{{The Hybrid spectral problem and Robin boundary conditions}}, \href{http://dx.doi.org/10.1088/0305-4470/38/21/017}{\emph{J. Phys. A} {\bfseries 38} (2005) 4735}, [\href{https://arxiv.org/abs/math/0409442}{{\ttfamily math/0409442}}].

\bibitem{Bianchi:2019sxz}
L.~Bianchi and M.~Lemos, \emph{{Superconformal surfaces in four dimensions}}, \href{http://dx.doi.org/10.1007/JHEP06(2020)056}{\emph{JHEP} {\bfseries 06} (2020) 056}, [\href{https://arxiv.org/abs/1911.05082}{{\ttfamily 1911.05082}}].

\bibitem{Giombi:2021uae}
S.~Giombi, E.~Helfenberger, Z.~Ji and H.~Khanchandani, \emph{{Monodromy defects from hyperbolic space}}, \href{http://dx.doi.org/10.1007/JHEP02(2022)041}{\emph{JHEP} {\bfseries 02} (2022) 041}, [\href{https://arxiv.org/abs/2102.11815}{{\ttfamily 2102.11815}}].

\bibitem{Giombi:2020rmc}
S.~Giombi and H.~Khanchandani, \emph{{CFT in AdS and boundary RG flows}}, \href{http://dx.doi.org/10.1007/JHEP11(2020)118}{\emph{JHEP} {\bfseries 11} (2020) 118}, [\href{https://arxiv.org/abs/2007.04955}{{\ttfamily 2007.04955}}].

\bibitem{Giombi:2021cnr}
S.~Giombi, E.~Helfenberger and H.~Khanchandani, \emph{{Fermions in AdS and Gross-Neveu BCFT}},  \href{https://arxiv.org/abs/2110.04268}{{\ttfamily 2110.04268}}.

\bibitem{Diatlyk:2024zkk}
O.~Diatlyk, H.~Khanchandani, F.~K. Popov and Y.~Wang, \emph{{Defect Fusion and Casimir Energy in Higher Dimensions}},  \href{https://arxiv.org/abs/2404.05815}{{\ttfamily 2404.05815}}.

\bibitem{eisenriegler1988surface}
E.~Eisenriegler and H.~Diehl, \emph{Surface critical behavior of tricritical systems}, {\emph{Physical Review B} {\bfseries 37} (1988) 5257}.

\bibitem{Prochazka:2019fah}
V.~Proch\'azka and A.~S\"oderberg, \emph{{Composite operators near the boundary}}, \href{http://dx.doi.org/10.1007/JHEP03(2020)114}{\emph{JHEP} {\bfseries 03} (2020) 114}, [\href{https://arxiv.org/abs/1912.07505}{{\ttfamily 1912.07505}}].

\bibitem{carslaw1910green}
H.~Carslaw, \emph{The green's function for a wedge of any angle, and other problems in the conduction of heat}, {\emph{Proceedings of the London Mathematical Society} {\bfseries 2} (1910) 365--374}.

\bibitem{book}
A.~Prudnikov, Y.~Brychkov and O.~Marichev, \emph{Integrals and Series. Vol. 2. Special Functions}, vol.~2.
\newblock 01, 1992.

\bibitem{Rich2018}
A.~Rich, P.~Scheibe and N.~Abbasi, \emph{Rule-based integration: An extensive system of symbolic integration rules}, \href{http://dx.doi.org/10.21105/joss.01073}{\emph{Journal of Open Source Software} {\bfseries 3} 1073}.

\end{thebibliography}\endgroup

\end{document}